\documentclass[preprint,12pt]{elsarticle}

\usepackage{lineno,hyperref}
\usepackage{xcolor}
\usepackage{lipsum}
\usepackage{blindtext}
\usepackage{longtable}
\usepackage{pifont}
\usepackage{caption}
\usepackage{subcaption}
\usepackage{lscape}
\usepackage{threeparttablex}
\usepackage{comment}
\usepackage{multirow}
\usepackage{xltabular}

\usepackage{etoolbox}
\usepackage{marginnote}

\newcommand{\modelconcept}[1]{{\fontfamily{phv}\selectfont#1}}
\newcommand{\challenge}[1]{{\textbf{#1}}}
\newcommand{\software}[1]{{\textit{#1}}}
\newcommand{\checkmark}{\ding{51}}
\newcommand{\xmark}{\ding{55}}
\newcommand{\replicationpkgurl}{\url{https://github.com/biazottoj/rp-sms-technical-debt-automation}}

\journal{Information and Software Technology}

\begin{document}

\begin{frontmatter}

\title{Technical Debt Management Automation: State of the Art and Future Perspectives}

\author[firstaddress,secondaddress]{João Paulo Biazotto\corref{mycorrespondingauthor}}
\cortext[mycorrespondingauthor]{Corresponding author}
\ead{j.p.biazotto@rug.nl}

\author[firstaddress]{Daniel Feitosa\corref{mycorrespondingauthor}}
\ead{d.feitosa@rug.nl}

\author[firstaddress]{Paris Avgeriou}
\ead{p.avgeriou@rug.nl}

\author[secondaddress]{Elisa Yumi Nakagawa}
\ead{elisa@icmc.usp.br}

\address[firstaddress]{Bernoulli Institute for Mathematics, Computer Science and Artificial Intelligence,\\ University of Groningen, The Netherlands}

\address[secondaddress]{Institute of Mathematics and Computational Sciences,\\ University of São Paulo, Brazil}

\begin{abstract}
\textbf{Context:} Technical debt (TD) refers to non-optimal decisions made in software projects that may lead to short-term benefits, but potentially harm the system’s maintenance in the long-term. Technical debt management (TDM) refers to a set of activities that are performed to handle TD, e.g., identification or measurement of TD. These activities typically entail tasks such as code and architectural analysis, which can be time-consuming if done manually. Thus, substantial research work has focused on automating TDM tasks (e.g., automatic identification of code smells). However, there is a lack of studies that summarize current approaches in TDM automation. This can hinder practitioners in selecting optimal automation strategies to efficiently manage TD. It can also prevent researchers from understanding the research landscape and addressing the research problems that matter the most. \\
\textbf{Objectives:} The main objective of this study is to provide an overview of the state of the art in TDM automation, analyzing the available tools, their use, and the challenges in automating TDM. \\
\textbf{Methods:} We conducted a systematic mapping study (SMS), following the guidelines proposed by Kitchenham et al. From an initial set of 1086 primary studies, 178 were selected to answer three research questions covering different facets of TDM automation. \\
\textbf{Results:} We found 121 automation artifacts that can be used to automate TDM activities. The artifacts were classified in 4 different types (i.e., tools, plugins, scripts, and bots); the inputs/outputs and interfaces were also collected and reported. Finally, a conceptual model is proposed that synthesizes the results and allows to discuss the current state of TDM automation and related challenges. \\
\textbf{Conclusion:} The research community has investigated to a large extent how to perform various TDM activities automatically, considering the number of studies and automation artifacts we identified. Nonetheless, more research is needed towards fully automated TDM, specially concerning the integration of the automation artifacts.
\end{abstract}


\end{frontmatter}

\section{Introduction}
\label{sec:intro}

Technical debt (TD) is a metaphor coined by Ward Cunningham~\cite{Cunningham1992} in 1992 and refers to poor decisions in software development that are made for several reasons, such as impending deadlines, budget constraints, and lack of knowledge~\cite{Rios2020}. When TD is not properly managed, it can accumulate to the point of harming the maintenance and evolution of software systems, leading to extra costs~\cite{Junior2022}.

According to Besker et al.~\cite{Besker2019}, developers spend around 23\% of their time dealing with TD. That amount of time usually corresponds to actions that would not be necessary if TD was not present, such as additional testing and refactoring. This subsequently can
increase
developers' workload 
and reduce their productivity, which can lead to higher costs in software development~\cite{Besker2019}. It also becomes a vicious circle: developers incur TD mostly because of time and budget restrictions~\cite{Rios2020}; these are the same reasons that a development team cannot afford expensive TDM activities to reduce their TD. 
In this context, the automation of TDM activities arises as a solution to reduce the overall effort of TDM, since it can decrease the load of developers in TDM activities and it can enable continuous TDM during any phase of the development~\cite{Khomyakov2019}.

Although some tools have been applied in the research and industrial communities (e.g., for identification~\cite{Martini2018-hy} and measurement~\cite{Amanatidis2020-da} of TD), we lack a clear overview of these tools within the set of TDM activities, as well as practical guidance on how to combine them and use them in contemporary software development~\cite{Junior2022}. Furthermore, to the best of our knowledge, there are no studies discussing the advantages and limitations of TDM automation, taking into account the different TDM activities and TD. We highlight that analyzing the technical aspects of such tools and their automation is crucial to improving integration in software development processes and realizing one main goal of applying them, i.e., to improve productivity by combining velocity with quality.

To address these shortcomings, we argue that it is necessary to report on existing approaches on TDM automation and 
derive potential improvements on them, which can further help to achieve higher levels of automation. To this end, we report on a systematic mapping study (SMS) that identified 121 automation artifacts\footnote{We use the term ``automation artifact'' to refer to any software used to automate TDM activities, i.e., tools, scripts, plugins, and bots. We highlight that, in our study, we do not consider datasets as automation artifacts.} used to support TDM automation and the associated TD activities and types. We also explore the characteristics of the automation artifacts, such as how and when to use them, and potential integration between them. Finally, we present several directions and challenges in TDM automation that can be explored in future work.

The rest of this paper is organized as follows. In Section~\ref{sec:background}, we provide the main TD-related terminology used in this study. In Section~\ref{sec:related_work}, we discuss related work and the main differences and contributions compared to our study. The research protocol, including objectives and research questions, are reported in Section~\ref{sec:design}. The results of this study are described in Section~\ref{sec:results} and the discussion about them in Section~\ref{sec:discussion}. The threats to validity and the efforts to mitigate them are written in Section~\ref{sec:tov}. Finally, the conclusions of this study are described in Section~\ref{sec:conclusion}.

\section{Background}
\label{sec:background}

In this section we present a summary of the main concepts related to TD and TD management. These concepts were mainly summarized Li et al.~\cite{Li2015}, in a previous study within the TD area.

The TD present in a system can be introduced in various moments of the software development process, since debt is not strictly related to the source code. Consequently, there are different TD types based on their source, e.g., source code, architectural decisions, tests, and infrastructure~\cite{Junior2022, Li2015, Alves2016}. Li et al.~\cite{Li2015} summarized nine types of TD:

\begin{description}
    \item[Requirements TD:] 
    relates to requirements elicitation and can refer to lack of requirements or misunderstanding of some of them, for instance;
    
    \item[Architectural TD:] bad decisions related to architectural design that could compromise 'internal' quality attributes of the software, such as evolvability or maintainability;
    
    \item[Design TD:] poor decisions made during the design phase, e.g., the division of responsibility among different classes; 
    
    \item[Code TD:] violations of code quality aspects, for instance, duplicate code, spaghetti code, duplicated variables, among others; 
    
    \item[Test TD:] non-optimal decisions taken in elaboration or execution of tests, e.g., lack of tests;
    
    \item[Build TD:] bad decisions that can harm the software building process, e.g., bad dependencies management; 
    
    \item[Documentation TD:] poor documentation in terms of correctness, completeness, and up-to-date aspects;
    
    \item[Infrastructure TD:] non-optimal decisions related to the selection of technologies for the software development, e.g., to use old technologies; and
    
    \item[Versioning TD:] problems in source code versioning, such as the lack of multi-version support. 
    
\end{description}

To keep TD under control, various activities have been proposed to help practitioners in managing TD~\cite{Alves2016, Guo2011, McGregor2012, Santos2013}. Li et al.~\cite{Li2015} summarized nine main activities that present in the literature. During \textbf{identification}, TD items are detected using several techniques, such as manual inspection or static code analysis. The identified items can be documented during the \textbf{representation and documentation} activities and stakeholders are informed about the TD items during the \textbf{communication} activity. The TD items can be then monitored during the \textbf{monitoring} activity, which ensures that unsolved TD items are under control. The \textbf{measurement} activity is used to quantify the amount of TD in a system, which in turn enables the \textbf{prioritization} activity, i.e., ranking TD items that must be solved first. 
TD items can then be fixed during the \textbf{repayment} activity, which also deals with the problems caused by TD accumulation. It is also possible to avoid undesired TD through the \textbf{prevention} activity.

In addition to the explanations of TD types and TD activities aimed at facilitating a comprehensive grasp of this research, it is pertinent to expound on the concept of self-admitted technical debt (SATD). SATD pertains to TD elements that developers themselves formally recognize as such~\cite{Maldonado2015}. For instance, this occurs when developers annotate source code with comments indicating discrepancies or areas in need of rectification. SATD frequently supplements other types of TD, since it imparts insights that alternative modes of TD identification might not unveil. To illustrate, opting for a suboptimal library is likely to be elucidated within a source code comment, whereas which would be challenging to identify only examining the source code (with static analysis tools, for instance)~\cite{Maldonado2015}.

Finally, it is also relevant to clarify what are ``replication package'' and ``supplementary material'', terms that were identified as relevant for our search string (see Section~\ref{sec:design} and Section~\ref{sec:pilot}). Those refers to material that are made available by researchers to improve the replicability of a scientific study. This packages can include figures, graphs, and data. Moreover, some tools and scripts developed by researchers are usually made available within a replication package, which is the reason to consider this terms in the search string.

\section{Related Work}
\label{sec:related_work}

In this section, we discuss other secondary studies that investigate tools that support TDM. To select this related work, we reviewed 29 secondary studies in the TD field, and analyzed them based on their research questions (RQs) and results. Studies closely related to ours, i.e., those that discuss TD tools and/or TDM automation, were selected, and are reported in this section. We also compare the studies with our own, and present a summary of this comparison in Table~\ref{tab:related_work}.

Li et al.~\cite{Li2015} conducted an SMS to investigate the current state of TDM, including what activities, tools, and management approaches have been used. They summarize 10 TD types, 8 activities, and 29 tools for TD management found in 94 primary studies. Regarding tools, the study points out some characteristics, such as their functionality, vendor, and TD types. The main similarity between this and our study is the analysis of the applicability of the tools. However, our study differs in that we focus on analyzing the automation artifacts and considering how they help in the automation of TDM as a whole. Moreover, we analyze other aspects of the automation artifacts, e.g., triggers and input/output.

In the study conducted by Khomyakov et al.~\cite{Khomyakov2019}, the authors discuss seven techniques and 10 tools for the measurement of TD, which they collected from 21 primary studies. They point out problems related to the complexity of the tools configuration and the limitations related to the support for different programming languages. We can summarize three main differences between their study and ours: first, their study just consider the measurement activity and does not cover the other eight activities as ours; second, although they consider three types of automation tools (tools, scripts, and plugins), they do not explore the differences among them nor how those differences can drive their usage; third, they do not explore the types of TD and the support for them, as we do in our study. Nonetheless, it is important to highlight the conclusion provided by the authors as an important motivation for our study, since they argue about the importance of TDM automation. 

Mumtaz et al.~\cite{Mumtaz2021} conducted an SMS to identify techniques and tools for architectural smells detection; architectural smells are considered one of the main types of architectural TD. The authors reviewed 85 primary studies and identified nine categories of techniques and 12 tools to support the techniques. Moreover, they discuss the types of smells identified by the tools, and also discuss some smells that do not have support yet (e.g., leaky encapsulation). The main similarity between the study of Mumtaz et al.'s and ours is the tools they identify, which we also cover in our study (all 12 tools). However, our study is much more inclusive in terms of TD types (they focused only on architectural TD) and TDM activities, since we consider all activities and they just focus on identification. Moreover, even though they analyze the tools' applicability, the scope of our work is different, because we focus on the automation aspects.

Lenarduzzi et al.~\cite{Lenarduzzi2021} analyze the current research on TD prioritization through an SMS. They selected 37 primary studies from which they collected the approaches and tools that are used, both by practitioners and researchers, to proceed with TD prioritization. Their main finding refers to the lack of consensus about the relevant factors to prioritize TD and how to measure those factors. Considering the TD types, they state that code and architectural debt are the most investigated types when considering how to prioritize TD items. The two main differences between their study and ours are: (i) the scope of their work is not focused on TDM automation; (ii) we do not focus only in one activity but cover all of them.

Avgeriou et al.~\cite{Avgeriou2021-fi} report a set of TD tools for TD measurement. In their study, the authors focus on three TD types (code, design, and architectural TD) and analyze the tools considering their features, popularity, and validation. Our study differs from theirs because we considered all TD types and TDM activities to collect the automation artifacts. Moreover, they report only 9 tools (namely CAST, Sonargraph, NDepend, SonarQube, DV8, Squore, CodeMRI, Code Inspector, and SymfonyInsight), and we report 121 automation artifacts (which include the 9 presented in their study), expanding the body of knowledge for more types of software that can be used for TDM automation. Moreover, our study discusses several different aspects related to the automation artifacts, such as their type of triggers, their input/output information and formats, and the possibility of integration among different automation artifacts.

Silva et al.~\cite{Silva2022} present an overview of 50 tools that can be used to manage TD, which they collected from 47 studies. There is an overlap between their work and ours since they also present results about what TD types and TD activities the tools support. However, our work differ in a number of ways. First, our scope considers not only tools but also any form of automation artifact (e.g., scripts, bots, plugins). Moreover, we have a different focus because we filter out automation artifacts that are not available or that do not automate a TD-related task. Combined with a longer study period (from 1992 to 2022, instead of from 2012 to 2019), we ultimately identified a higher number of automation artifacts (121 automation artifacts from 178 primary studies). Second, our research objectives are broader since we explore more characteristics of the automation artifacts, considering their inputs/outputs and the possible integration among the automation artifacts. We also provide an analysis of the usage of the automation artifacts, considering their types, triggers, and standalone execution. Finally, our discussion about the tools is not similar. Silva et al. focus on how tools have been supporting the TDM in terms of features, while we analyze the automation artifacts to understand how they can be used to automate the TDM in a development environment. 

To the best of our knowledge, no previous study has investigated the artifacts considering the aspects that we highlighted in the previous paragraphs (e.g., in terms of inputs/outputs and possible integrations). Moreover, we could not find works that discuss not only artifacts' features but also how they could be used and integrated within development activities (e.g., coding) and the development environment (e.g., integrated into a continuous integration manager). This lack of information about artifacts' usage can negatively impact practitioners when they are choosing which artifacts to use to support TDM. Thus, our study fills this gap and expands the state of the art by discussing how to support TDM activities using automation artifacts and possible workflows using those artifacts.

To facilitate the comparison of our study and the related works, we use the set of criteria listed in Table~\ref{tab:related_work}. In summary, our work provides the following unique contributions compared to the rest of related work:

\begin{itemize}
    \item \textbf{Analysis and Discussion of TDM Automation:} to the best of our knowledge, our study is the first one that considers all TD types and TDM activities to discuss the current TDM automation approaches, including their strengths and limitations. Moreover, we also provide a discussion about the challenges and the future directions of this topic, which is not present in any other study;
    
    \item \textbf{Discussion about the usage of the automation artifacts:} our study discusses the usage of the automation artifacts considering their application in a development environment. In this context, we provide an analysis of the integration between the automation artifacts and between automation artifacts and other tools (e.g., IDE). Besides, we analyze how this integration improves the level of automation of TDM; and
    
    \item \textbf{Analysis of technical aspects of the automation artifacts: } related works only discuss the features of the automation artifacts, whereas this study expands on other perspectives, such as the triggers that initiate the automation and the inputs/outputs of the automation artifacts.
\end{itemize}

\begin{table}[ht]
    \centering
    \begin{threeparttable}
    \caption{Comparison with related works}\label{tab:related_work}
    \begin{tabular}{p{2.6cm}
                 p{1.0cm}
                 p{0.8cm}
                 p{1.2cm}
                 p{2cm}
                 p{0.7cm}
                 p{0.7cm}
                 p{0.8cm}
                 p{0.7cm}
                 p{0.7cm}}
\\
\hline
\textbf{reference} & \textbf{method} &\textbf{last year*} & \textbf{\# studies} & \textbf{\# automation artifacts} & \textbf{TDT} & \textbf{TDA} & \textbf{TDMA} & \textbf{TA} & \textbf{INT} \\\hline

Li et al.~\cite{Li2015}                 & SLR   & 2013 & 94  & 29     & \checkmark & \checkmark & \xmark     & \xmark     & \xmark      \\
Khomyakov et al.~\cite{Khomyakov2019}   & SLR   & 2017 & 21  & 10     & \xmark     & \checkmark & \xmark     & \xmark     & \xmark      \\
Mumtaz et al.~\cite{Mumtaz2021}         & SMS   & 2019 & 85  & 12     & \xmark     & \xmark     & \xmark     & \xmark     & \xmark      \\
Lenarduzzi et al.~\cite{Lenarduzzi2021} & SMS   & 2019 & 37  & 12     & \checkmark & \xmark     & \xmark     & \xmark     & \xmark      \\
Avgeriou et al.~\cite{Avgeriou2021-fi}  & Mixed & 2019 & 50  & 9      & \checkmark & \xmark     & \xmark     & \xmark     & \xmark      \\
Silva et al.~\cite{Silva2022}           & SMS   & 2019 & 50  & 47     & \checkmark & \checkmark & \xmark     & \xmark     & \xmark       \\
Our study                               & SMS   & 2022 & 178 & 121    & \checkmark & \checkmark & \checkmark & \checkmark & \checkmark  \\
\hline
\end{tabular}
\begin{tablenotes}
    \footnotesize
    \item[] \textbf{TDT}: The study discusses all TD types presented in Section~\ref{sec:background}.
    \item[] \textbf{TDA}: The study discusses all TDM activities presented in Section~\ref{sec:background}.
    \item[] \textbf{TDMA}: The study discusses TDM automation.
    \item[] \textbf{TA}: The study discusses technical aspects of automation artifacts (e.g., provided triggers and interfaces).
    \item[] \textbf{INT}: The study discusses integrations between automation artifacts
    \item[] *last year included in the study collection.
\end{tablenotes}
\end{threeparttable}
\end{table}

\section{Study Design} 
\label{sec:design}

This section describes the design of our SMS, followong the guidelines of Kitchenham et al.~\cite{Kitchenham2015}. 

\subsection{Research Objective}
\label{sec:objective}
To define the objective of our study, we adopted the structured goal description as defined in the Goal-Question-Metric (GQM) approach~\cite{vanSolingen2002}. Specifically, the objective of our study is to \textit{``\textbf{analyze} primary studies from the software engineering literature \textbf{for the purpose of} understanding the state of the art in the automation of TDM \textbf{with respect to} the automation artifacts, the related TDM activities and types, and their usage in the software development process \textbf{from the point of view of} researchers and practitioners \textbf{in the context of} TD.''} 

\subsection{Research Questions}
The research questions (RQ) are derived from the study's objective and are the basis for creating the inclusion and exclusion criteria and the extraction form (see Sections~\ref{sec:sd_selection} and~\ref{sec:sd_extraction}). 

\vspace{6pt}
\noindent \textbf{RQ1 - What automation artifacts are available for the automation of TD management?}
\begin{description}
    \item[RQ1.1 -] What is the type of software of the automation artifact (e.g., script, tool, plugin)?
    \item[RQ1.2 -] What are the inputs and outputs of the automation artifact?
    \item[RQ1.3 -] Which is the evidence level of the software automation artifact?
\end{description}

Different automation artifacts can automate TDM activities, such as \textit{tools}~\cite{Avgeriou2021-fi} and bots~\cite{Phaithoon_2021}. Hence, the objective of this RQ is to catalog the available automation artifacts that can perform TDM-related activities. We further refine this RQ into three sub-questions. In RQ1.1, we intend to understand the automation artifacts' type (e.g., script, tools) and their characteristics (e.g., stand-alone execution).
RQ 1.2 aims to catalog the associated inputs and outputs of the automation artifacts. Finally, in RQ1.3, we aim to review the automation artifacts' evidence level, considering the classification presented by Alves et al.~\cite{Alves2010_evidence_level}, which takes into account the type of evidence provided by the primary studies (e.g., academic studies and industrial studies). Thus, RQ1.3 provides an overview of the validation of the automation artifacts. The knowledge provided by RQ1 can help practitioners become aware of the current range of automation artifacts and choose those that fit their own needs.

\vspace{6pt}
\noindent \textbf{RQ2 - What TDM activities and TD Types are supported by the automation artifacts?}

Although related work has already cataloged the automation of some TDM activities, especially regarding tools~\cite{Li2015, Silva2022}, it is also relevant to investigate 
the TDM automation considering all TDM activities and TD types, specially taking into account different types of automation artifacts (e.g., scripts and plugins). 
For both TDM activities and TD types, we consider the lists 
presented by Li et al.~\cite{Li2015} (see Section \ref{sec:background}). We note that although this classification was proposed in 2015, it is still contemporary and widely present in the TD literature~\cite{Junior2022}. 
Answering this RQ can help practitioners and researchers 
to select the optimal tools for the task at hand, or the problem to be solved. Moreover, this RQ reveals the current support for TD types and TDM activities, which can also drive future research by exposing which types and activities are not yet well supported.

\vspace{6pt}
\noindent \textbf{RQ3 - How are automation artifacts used during software development?}
\begin{description}
    \item[RQ3.1] What triggers are used to initiate the automation process?
    \item[RQ3.2] Can the automation artifacts be integrated?
\end{description}

This research question aims at understanding the use of the automation artifacts in the automation of TDM activities along two dimensions, that correspond to the two sub-questions. RQ3.1 aims at understanding the events that are responsible for initiating the automation process, also looking into the level of human intervention that is necessary. 
RQ3.2 aims at understanding the possibility of integration between different TDM automation artifacts as, well as, integration between automation artifacts and other tools (e.g., as plugins in an IDE). This RQ can help practitioners on understanding how to incorporate the automation artifacts within their software development practices and tools.

\subsection{Search Strategy}
\label{sec:strategy}
To conduct this study, we choose a search strategy similar to the one used by Junior et al.~\cite{Junior2022}. First, we perform an automatic search using the Scopus Database\footnote{\url{https://www.scopus.com/}}, which indexes more than 7,000 publishers~\footnote{\url{https://www.elsevier.com/solutions/scopus/how-scopus-works}}, including the most relevant for the field of Software Engineering. Second, to compensate the use of a single database (albeit a comprehensive one), we perform snowballing~\cite{Wohlin2014}. In particular, we consider both forward snowballing (i.e., search citations to primary studies) and backward snowballing (i.e., search primary studies' citations). We note that  our decision is aligned with recent literature showing that using a hybrid strategy provides better results when compared to using only automated search~\cite{Mourao2020, Wohlin2022,Fernandez2022}.

To construct our search string, we take inspiration from the work of Li at.al~\cite{Li2015}; they used ``debt'' as a search string since ``technical debt'' is not always captured in studies' title, abstract, and keywords, especially when the primary study is related to a specific type of TD (and then authors use terms like ``architectural debt'')~\cite{Li2015}. Moreover, using ``debt'' as a search string provides a very inclusive search, allowing us to analyze a higher number of studies. To filter the recovered studies and calibrate the search string to the scope of our study, we combined the keyword ``\textit{debt}'' with several keywords related to ``\textit{automation and tooling}'' (e.g., ``tool'', ``automation'', and ``bot''). To calibrate our search string, we performed several rounds of automatic search using the Scopus database and tested different combinations of keywords. In each round, the first two authors manually checked the number of studies and their relevance to answering the RQs. The set of returned studies stabilized on the fourth round when we proceeded to conduct a pilot study to evaluate the data extraction and synthesis (see Section~\ref{sec:pilot}).

Regarding the range of years, we reused the set of studies reported by Li et al.~\cite{Li2015} (which searched from 1992 to 2013) and extended our set with studies published between 2013 and 2022. Our rationale to reuse Li et al.'s set of studies is two-fold: (a) the search string provided by them (i.e., ``debt'') is inclusive, covering possibly all studies related to TD published until 2013; and (b) their work also covers tooling for TDM (specifically in RQ8); thus, the scope of our work is included within the scope of their work.

Also, we restrict the results to the Computer Science field, since ``debt'' is a keyword commonly used in other fields (e.g., financial debt). The final search string is defined as follows: \textit{title-abstract-keywords(``debt'') AND full-text(``script*'' OR ``replication package'' OR ``supplementary material'' OR ``tool*'' OR ``artifact'' OR ``automat*'' OR ``artefact'' OR ``plugin'')}.

\subsection{Study Selection}
\label{sec:sd_selection}

In the selection phase, we apply a set of inclusion and exclusion criteria throughout a series of selection rounds. The criteria and selection rounds are described in the following.

\subsubsection{Inclusion and Exclusion Criteria}

The following inclusion (IC) and exclusion (EC) criteria were used in this study:

\begin{description}
    \item[IC1] - The primary study proposes, uses, or points to an automation artifact that automates a task within at least one TD management activity.
    \item[EC1] - The primary study 
    is not related to software engineering;
    \item[EC2] - The study shows only the design of an automation artifact;
    \item[EC3] - The study does not provide a URL to the automation artifact;
    \item[EC4] - The full text of the primary study is not available;
    \item[EC5] - The primary study is a table of contents, short course description, or summary of a conference/workshop;
    \item[EC6] - The primary study is written in a language other than English; and
    \item[EC7] - The study is a previous version of a more complete one that also mentions the same automation artifact.
\end{description}

These ICs and ECs are based on the research objective and RQs. Since our main goal is to map the available automation artifacts for TDM, IC1 includes all studies pointing to an automation artifact. In the same direction, EC1 removes studies not related to software engineering since the term ``debt'' can be used in other fields (e.g., financial debt).

As our focus is the automation artifacts that could be used in practice, EC2 removes studies that present just a design of an artifact, since it would not be possible to analyze some technical aspects of such artifacts (e.g., inputs/outputs and triggers). Regarding EC3, we needed the URL pointer to the actual automation artifact to be able to analyze it. Alternatively, we could search for the artifact online using its name, but this is not feasible given the number of studies we are filtering and the risk of finding the wrong artifact. Moreover, it is not unreasonable to expect that an artifact deemed relevant by the authors would be properly referenced (i.e., through a URL to a replication package, homepage, repository or similar). The remaining criteria (EC4-EC7) are general and are based on guidelines and previous works within software engineering and TD field~\cite{Junior2022,Alves2016,Kitchenham2015,Wohlin2022}.

\subsubsection{Selection Process}

The study selection comprises the following rounds: in the \textbf{first round}, the authors filtered papers by reading \textbf{title, abstract, keywords, and venues}, and applied the criteria IC1, EC1, EC5 and EC6. EC2, EC3, EC4, and EC7 cannot be applied in this round since they cannot be checked based on the metadata. In the \textbf{second round}, the authors downloaded all available studies and filtered them based on the \textbf{Introduction and Conclusion} sections and applied EC4-EC6. In this round, the authors also applied EC2 and EC3, to check if the study shows an implemented automation artifact (i.e., that can be downloaded and executed). We note that if an automation artifact is referenced via another paper, we followed that paper (and recursive references if needed) in search of a URL.

\begin{figure}[ht]
    \centering
    \includegraphics[width=\textwidth]{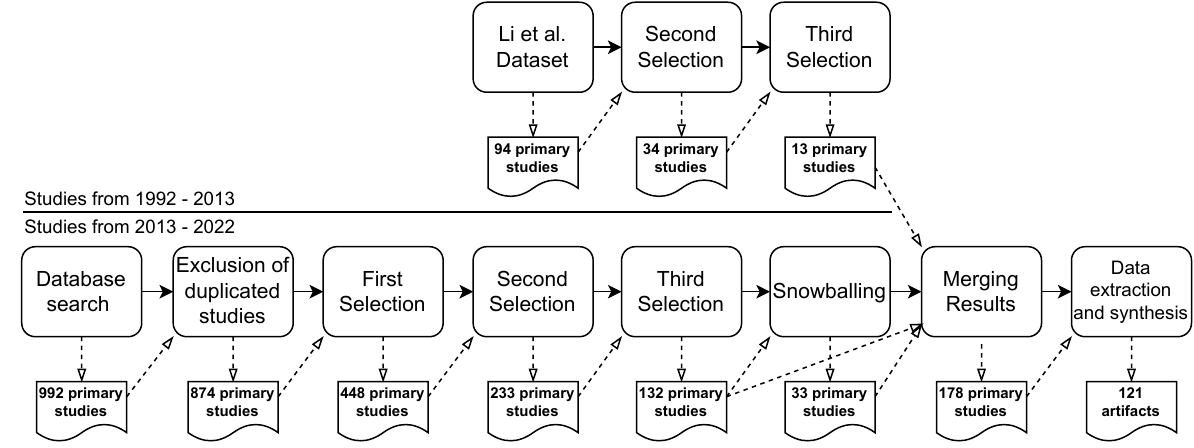}
    \caption{Selection process}
    \label{fig:selection_process}
\end{figure}

Finally, in the \textbf{third round}, the authors filtered the studies based on the \textbf{full text}, applying EC7. The first search in the Scopus Database was on June 30th 2022.  
Figure~\ref{fig:selection_process} summarizes the process we followed and the results that were obtained after each step. For the dataset of Li et al.~\cite{Li2015}, we followed the second and third round of selection, since the first round of selection (i.e., title, abstract, keywords, and venues) was already done in the original study.

\subsection{Data Extraction, Analysis, and Synthesis}
\label{sec:sd_extraction}

The main goal of this study is to characterize the current research on TDM automation by analyzing information about the automation artifacts that are used during the automation process. To achieve this goal, we built two datasets: one with information related to the studies (from where we collected the automation artifacts) and a second with details on the automation artifacts. The first is described in Table~\ref{tab:extraction_form} and presents the data extracted from each study, which were used for demographic purposes (see Section~\ref{sec:results_demographic}). The second is presented in Table~\ref{tab:extraction_form_automation artifacts} and shows the data extracted regarding each automation artifact. It is important to highlight that some data were retrieved from the automation artifact's documentation, such as the input/output formats. The RQs that were answered using that information are also presented.

\begin{table}
    \centering
    \small
    \caption{Data extracted from each study}
    \label{tab:extraction_form}
    \begin{tabular}{p{3.0cm}p{9.7cm}}
\\
\hline    
\textbf{variable name} & \textbf{description} \\ \hline 

article\_title & The title of the study\\

author & Name of the authors\\

author\_type & Type of affiliation of the authors (Academia, Industry, or Both)\\

year & The publication year of the study\\

venue & The venue were the study was published\\

venue\_type & Type of venue where the study was published (Journal, Conference, Book Chapter)\\

doi & DOI of the study\\

reference & Bibtex entry of the study\\

objective & A short summary of the study's objective\\

automation artifacts & The list of automation artifacts pointed out in the paper\\
\hline
\end{tabular}
\end{table}

\begin{table}
    \centering
    \small
    \caption{Data extracted from each automation artifact}
    \label{tab:extraction_form_automation artifacts}
    \begin{tabular}{p{3.5cm}p{7cm}p{2.1cm}}
\\
\hline    
\textbf{variable name} & \textbf{description} & \textbf{results} \\
\hline

studies\_citing\_artifact & ID of the studies that mention the artifact; this variable relates artifacts to studies & Demographic info.\\

name & The name of the automation artifacts & Demographic info.\\ 

link & The link to the automation artifact & Demographic info.\\ 

type & The software type of each automation artifact (e.g., plugin) & RQ1.1\\ 

run\_standalone & Whether it is possible to run the automation artifacts as standalone & RQ1.1 \\ 

input\_info & The information used by each automation artifact to automate the task (e.g., source code) & RQ1.2\\ 

input\_fmt & The format of the information used by the automation artifact (e.g., XML) & RQ1.2, RQ3.2   \\

output\_info & The information provided by the automation artifact after the automation process & RQ1.2 \\ 

output\_fmt & The format of the information provided by the automation artifact (e.g., XML) & RQ1.2, RQ3.2  \\

evidence & The type of evidence we found in the literature (e.g., industrial studies) & RQ1.3\\ 

td\_type & The types of TD that are supported by the automation artifact & RQ2  \\ 

tdma & The TDM activities that are supported by the automation artifact & RQ2   \\ 

trigger & The type of trigger that starts the execution of the automation artifacts (e.g., human trigger) & RQ3.1  \\ 

interface-type & The type of interface provided by the automation artifact (e.g., api) & RQ3.2 \\
 
is\_integrated & Whether we found evidence of integration among the automation artifacts and other automation artifacts & RQ3.2  \\ 

can\_integrated & Whether the automation artifact could be integrated with other automation artifacts, considering the input/output format and information,  & RQ3.2  \\

\hline
\end{tabular}
\end{table}

The data analysis and synthesis of our study is based on the constant comparison method~\cite{Corbin2008, Stol2016} for qualitative data and frequency analysis for quantitative data. To answer \textbf{RQ1} (automation artifacts available), we used the information related to characteristics of the automation artifacts, i.e., the data items \textit{type, run\_standalone, input\_info, input\_fmt, output\_info, output\_fmt}, and \textit{evidence}, as described in Table~\ref{tab:extraction_form}. For \textbf{RQ2}, we focused on understanding which TD Types and Activities are supported by the automation artifacts. In this context the items \textit{td\_type} and \textit{tdma} were used, providing an overview of the current state to the automation of TDM. For \textbf{RQ3}, we aimed at understanding the usage of the automation artifacts, considering their triggers, 
and the integration among the automation artifacts. To answer this question, we used, \textit{trigger}, \textit{interface}, \textit{is\_integrated}, and \textit{can\_integrated}.

\subsection{Pilot Study}
\label{sec:pilot}
We conducted a pilot study to validate and further improve our study design.
For that, we applied the search string (``debt'') on Scopus and recovered around 2,100 results, from which we randomly selected 50 studies. We conducted the three rounds of selection and the data extraction on these studies (see Sections~\ref{sec:sd_selection}~and~\ref{sec:sd_extraction}). Out of the 50 initial studies in the sample, 13 proposed or pointed to a automation artifact that automatically performs a TDM task.

Based on the pilot, we validated and calibrated our search string, the inclusion and exclusion criteria, and the extraction form.
One example of the search string calibration is the addition of the terms \textit{``replication package''} and \textit{``supplementary material''}. During the pilot study, we realized that sometimes the authors make their automation artifacts available through replication packages (e.g., when a script was developed). Thus, after the pilot, we added those new terms to the search string.
Finally, the pilot also provided directions for conducting this study, allowed to validate which type of information can be used to answer the RQs and ensuring that this study is relevant to the TD community.

Considering both the pilot study and the related work, we were confident that conducting this SMS could provide directions to the TD community, cataloging existing tools and uncovering promising research directions.

\section{Results}
\label{sec:results}
In this section, we present the results of our SMS. Specifically, Section~\ref{sec:results_demographic} explores demographics regarding the distribution of studies over the years, the main venues and author affiliations. The answers to our three RQs are presented in Sections~\ref{sec:results_sw_artifcats}~to~\ref{sec:results_usage_automation artifacts}, where we respectively discuss the available automation artifacts, their scope and how they are used within TDM activities.

\subsection{Demographic Information}
\label{sec:results_demographic}

\begin{figure}[ht]
     \centering
     \begin{subfigure}[b]{\textwidth}
         \centering
         \includegraphics[width=\textwidth]{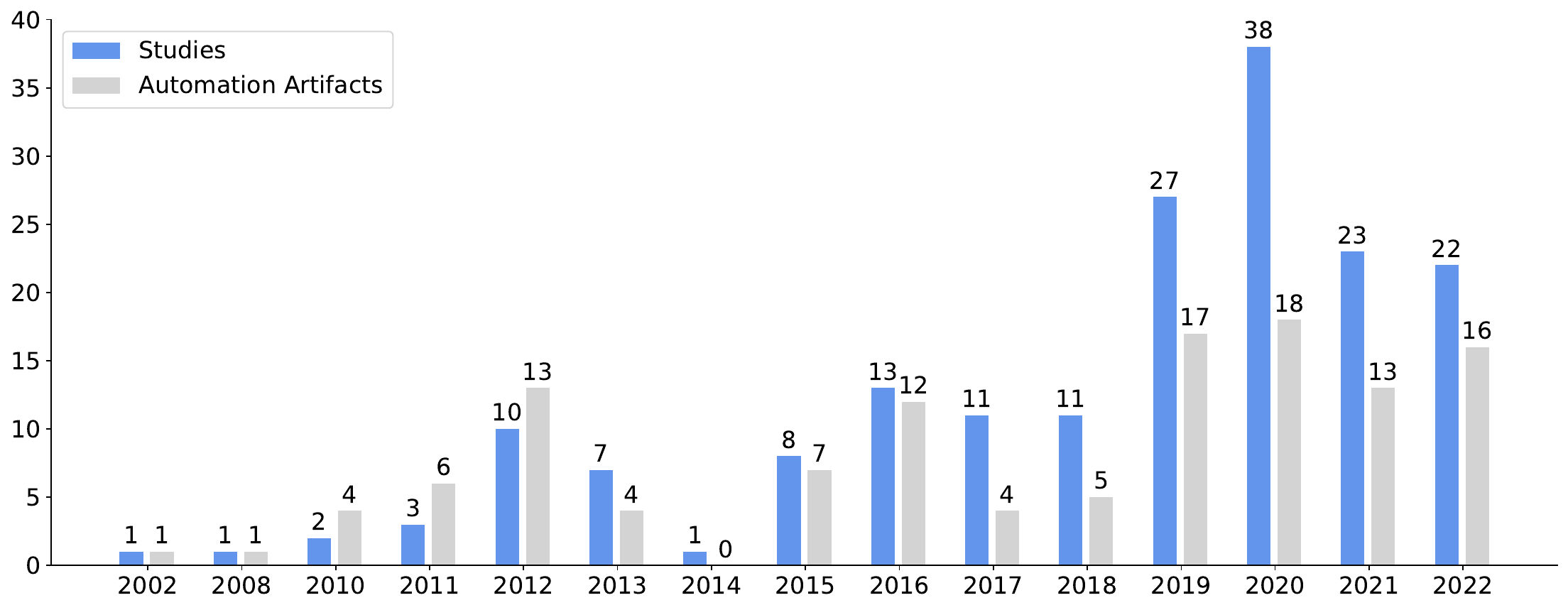}
         \caption{Distribution per year}
         \label{fig:studies_year}
     \end{subfigure}
     \begin{subfigure}[b]{3.8cm}
         \includegraphics[width=\textwidth]{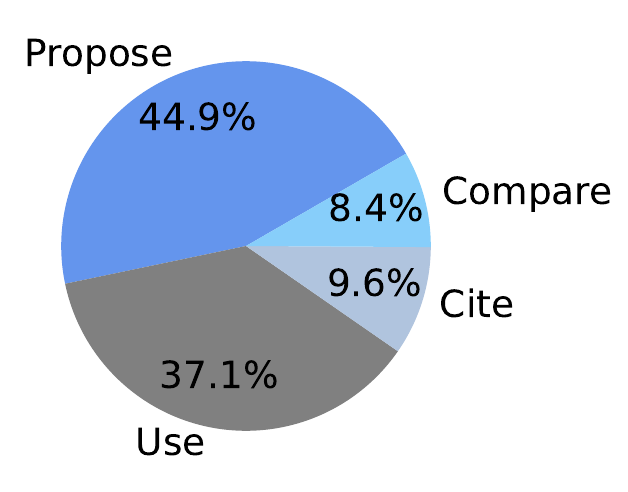}
         \caption{Distribution of studies based on the artifact role}
         \label{fig:study-type}
     \end{subfigure}
     \hfill
     \begin{subfigure}[b]{4cm}
         \includegraphics[width=\textwidth]{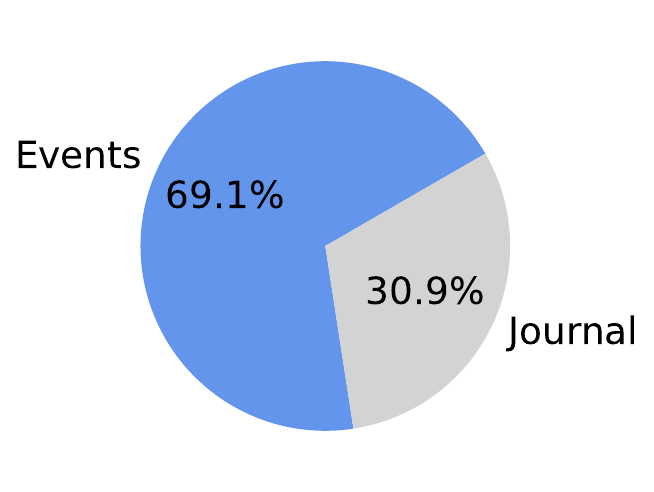}
         \caption{Distribution of studies per type of venue}
         \label{fig:studies_venue_type}
     \end{subfigure}
     \hfill
     \begin{subfigure}[b]{4.7cm}
         \includegraphics[width=\textwidth]{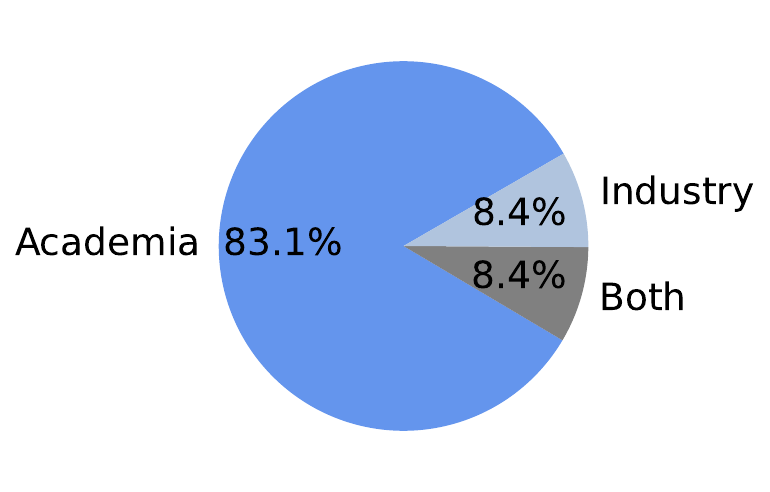}
         \caption{Distribution of studies per author affiliation}
         \label{fig:studies_author_type}
     \end{subfigure}
        \caption{Demographic information of the studies and automation artifacts}
        \label{fig:demographic}
\end{figure}

After the selection steps, we extracted data from 178 studies. The selection process covered the time span between 1992 and June/2022. However, the first study we found that pointed to an automation artifact was published in 2002, as shown in Figure~\ref{fig:studies_year}, which illustrates the studies we identified, organized by year. Moreover, Figure~\ref{fig:studies_year} also presents how many unique automation artifacts were cited for the first time in a certain year. We note  that we collected studies published until June/2022, when the first search was performed. Thus, studies published after this month are not covered. Considering the data we had available, it is evident that the number of studies and automation artifacts is growing over the years.  
Of all the selected studies, at least 67\%~(121/178)\footnote{This percentage might be higher considering we did not collect data from the second half of 2022.} of them were published in the last five years, which shows that the investigation of the use of automation artifacts to support TDM is a ``hot'' topic.

Figure~\ref{fig:study-type} shows the distribution of studies based on whether the artifact is \textit{proposed}, \textit{compared}, \textit{used}, or just \textit{cited} in the study. Around 45\% (80/178) of the studies \textit{propose} an artifact, i.e., introduce the artifact for the first time and present its main characteristics and functionality. While this may indicate that the community is interested in finding new solutions for automation in TDM, it may also indicate that some artifacts are still immature and demand evaluation in industry and open-source contexts. Next, around 8\% (15/178) of the studies \textit{compare} two or more tools, e.g., presenting the best tool at performing a certain task such as identifying TD~\cite{Zazworka_2013}. We also see that 37\% (66/178) \textit{use} an artifact either in a TDM approach (e.g., in the industry) or in empirical studies that investigate TD. Finally, the remaining 10\% (17/178) of the studies only cite an artifact as an example of a possible solution for TDM but do not present deeper information about its functionality or usage.

Regarding the venues where we found the studies, according to Figure~\ref{fig:studies_venue_type}, 69\%~(123/178) of the studies were published in events (conferences, symposiums, and workshops) and  31\%~(55/178) were published in journals. The International Conference on Technical Debt (TechDebt) was the most common venue in our dataset with 22/178 studies. Regarding journals, the Journal of Systems and Software (JSS) was the most common (8/178). 

To measure the level of industrial involvement in the development and use of automation artifacts for TDM automation, we extracted data about the authors' affiliation, presented in Figure~\ref{fig:studies_author_type}. We found that 148/178 studies were conducted only by academic authors, while 15/178 involved authors only from the industry. Finally, in 15/178 studies an academia/industry collaboration was shown. This indicates that, although industry has substantial interest in TD~\cite{Rios2020, Besker2019, Martini2018-hy}, there is not much research on TDM automation published from industry, since only 17\% (30/178) of the studies have an author affiliated with companies.

\vspace{6pt}
\noindent\fbox{%
    \parbox{\textwidth}{%
    \textbf{\textit{Finding 1:}} Around 45\% of the studies \textit{propose} an artifact, which may suggest that the community is actively finding new solutions for TDM automation. Also, most of the studies were published in the last five years (124/178), indicating that this topic is of increasing interest to the TD community.
    Moreover, academia is responsible for around 83\% of the published research work on TDM automation, while the participation from industry is rather limited.
    }%
}

\subsection{Automation artifacts for TDM automation}
\label{sec:results_sw_artifcats}

In \textbf{RQ1}, we analyzed the existing automation artifacts for the TDM automation. In total, we identified 121 automation artifacts that can be used in the automation of TDM activities and are available online, i.e., they can be installed by practitioners and researchers. Due to space constraints, the list of all automation artifacts is omitted here but can be found in our replication package\footnote{\replicationpkgurl}.

\subsubsection{Types of automation artifacts}
\label{sec:results_sw_artifcats_types}
Considering the \textbf{automation artifacts type}~(RQ1.1), we identify four different types according to their characteristics of deployment and execution (among others, if they run standalone, if there is the possibility of configuration, and how much human intervention is required to execute them).

\textbf{Tools} are the most common type of automation artifacts, with approx. 47\%~(57/121) of them classified into this type. This term has been used to refer to automation artifacts by many primary studies
~\cite{Amanatidis2020-da, Avgeriou2021-fi, Lenarduzzi2020-vw, Fontana2016-ag}. 
Automation artifacts introduced as tools have three main characteristics.
First, tools do not depend on other automation artifacts to be executed. Second, they provide usage interfaces, i.e., their functionalities are well-organized,
and easy to configure and execute. Those interfaces can be Graphical User Interface (GUI), command line, or APIs, which can be used by other automation artifacts or with customized scripts. Third, tools are highly configurable i.e., they implement several types of metrics, rules, and analyses that can be used (or not) by the practitioners, considering their context.  SonarQube\footnote{\url{https://www.sonarqube.org/}} is an example of a tool. The calculation of code metrics and amount of TD, for example, can be accessed through a GUI or collected by APIs to be used in diverse types of analysis. 
Arcan\footnote{\url{https://essere.disco.unimib.it/wiki/arcan/}} is another example of a tool, which can identify architectural smells (e.g., cyclic dependency), a type of architectural TD, and provide the results through several formats (e.g., csv).
A third example is Designite\footnote{\url{https://www.designite-tools.com/}}, a tool that can deal with design TD, focusing on identifying design smells; it also provides dashboards to analyze the analysis results.
The main advantages of using tools are the flexibility of configuration and the wide range of different information provided.  However, some effort is needed to configure the tools in the development life cycle, e.g., integrating and configuring them in the continuous integration (CI) pipeline; this can be a disadvantage of using tools when compared with other types of automation artifacts that are executed more easily (e.g., scripts).

\textbf{Plugins} represent around 32\%~(38/121) of the automation artifacts. Several studies have presented this term to refer for automation artifacts, such as Masmali et al.~\cite{Masmali2021-ps}, which presents jDeodorant as plugin for Eclipse IDE. The main characteristic of plugins is that they do not run standalone but are installed within an IDE or in other automation artifacts (e.g., tools). They are responsible for adding a new capability to an existing automation artifact; they are installed on top of other automation artifacts, so it is not necessary run their source code directly (as is the case for scripts).
For instance, jDeodorant\footnote{\url{https://github.com/tsantalis/JDeodorant}}, a plugin developed for Eclipse IDE\footnote{\url{https://eclipseide.org/}}, makes it possible to identify code smells, in real-time, during the coding process. Regarding this plugin, Paiva et al.~\cite{Paiva2017} presented an empirical study that compared it with PMD and JSpirt. Their results show that the precision of JDeodorant is higher than the other ones, i.e., it is able to identify more and correct smells. However, it presents a high number of false positives.

It is also important to highlight that several tools are also available as plugins, with fewer configuration options and specific functions. Specifically, 22/38 plugins are simpler versions of tools. For instance, Checkstyle\footnote{\url{https://checkstyle.sourceforge.io/}} is available both as a tool (that can be run standalone) and as a plugin for several IDEs (such as Eclipse IDE and IntelliJ\footnote{\url{https://www.jetbrains.com/idea/}}). We observed two main differences between plugins and tools: (i) plugins provide real-time feedback directly in the IDE, which can aid developers with suggestions to directly modify the source code; and (ii) tools
provide a deeper understanding of the quality of the code base, giving more decision-making information about the TD items; for instance, the plugin provided by DesigniteJava can not generate dashboards directly, but only through the tool version of DesigniteJava.

Around 18\%~(22/121) of the automation artifacts are classified as \textbf{scripts}, which refer to pieces of code executed to automate some task. Usually, scripts use results from other automation artifacts (especially tools) to perform new tasks with that information. An example of study that uses this term to refers to an automation artifact is Sas et al.~\cite{Sas_2019}, which presents a script (piece of source code) to collect results from Arcan and track them in source code. Scripts are designed and implemented in a specific context,
it is complex to reuse them in another context (e.g., a script that collects data generated by SonarQube can only be used in this context). Two examples are \textit{debtfree}\footnote{\url{https://github.com/HuyTu7/DebtFree}} and \textit{piranha}\footnote{\url{https://github.com/uber/piranha}}. The first is a script that implements machine learning algorithms to identify SATD. It can execute a single task (label comments), considering a specific input of pre-processed code comments. The second, \textit{piranha}, is used to automate the repayment of a specific type of code debt in Java programs, i.e., flag features. These flags encapsulate 
parts of the code that are still under development, and are not ready to be used yet. Nonetheless, developers can forget to remove the flags, leading to maintenance issues. Considering the name and types of the flags, \textit{piranha} can refactor the source code, sending the results as pull requests in GitHub. The main advantage of scripts is related to their easiness of configuration and execution. However, since they are developed for a specific context, they are not configurable, which harms their flexibility. This may be an explanation for the lack of studies that compare two or more scripts in terms of functionalities and usage in a TDM approach.

Another type of automation artifact we identified is \textbf{bots}, which correspond to nearly 3\%(4/121) of the automation artifacts. They are similar to scripts; however, they run in the background, i.e., they are triggered automatically when some actions happen. Phaithoon et al.~\cite{Phaithoon_2021}, for instance, used the term bot to introduce FixMe\footnote{\url{https://www.fixmebot.app/}}.
Fixme can identify SATD in code comments and runs on top of the Probot Framework, which enables its integration with GitHub, a git platform for code management. Every time source code is committed to GitHub, FixMe analyses it and generates several issues, each corresponding to an instance of detected SATD. Finally, FixMe keeps a database about TD items, monitoring them. Although we just identified a few examples of bots, they present an interesting characteristic that can help practitioners in the automation of TDM: they are automatically triggered. With bots, for example, it could be possible to automate the documentation of TD, reducing the necessity of human intervention. However, we did not identify studies that analyze this characteristic in a TDM approach. Besides, no study compares the usage of different bots, which poses the necessity of more studies to understand which bot would be more effective in a certain scenario.

From all examined automation artifacts, tools seem to be the most preferred by practitioners/researchers to deal with TDM activities and tasks. Presumably, the characteristics of standalone deployment and flexibility are perceived as a way to improve the efficiency of automation. For example, Sharma et al.~\cite{Sharma2016-ca} discuss their choice of using tools with different types of rules, in which they can define what rules will be used. Moreover, some tools support the installation of plugins, 
which provide new functionalities to the tool, thus supporting higher levels of automation.
From another perspective, bots can also be used in future automation approaches, considering their integration in the development workflow. Since they do not need human intervention when being executed, they seem to be an alternative to introducing TDM automation in the development life-cycle, relieving developers from tedious tasks. However, we found only little evidence about bots, so we argue that a deeper investigation into their use is still needed. It is also necessary to compare the four types of artifacts to understand both developers' preferences and also the performance of different artifacts when performing a specific TD activity.

\vspace{6pt}
\noindent\fbox{%
    \parbox{\textwidth}{%
    \textbf{\textit{Finding 2:}} Four different types of automation artifacts could be used for TDMA automation. Among them, tools are the most preferred type, especially for their standalone deployment and the range of different configurations. In addition, bots arise as an alternative for the automation of some tasks without human intervention, although currently there is little evidence on their use.
    }%
}

\subsubsection{Automation artifacts' inputs and outputs}
\label{sec:results_sw_artifcats_inputs}

To answer \textbf{RQ1.2}, we extracted two pieces of information regarding the input used in the 121 automation artifacts we identified. First, we analyzed the~\textbf{information} that is used as input by the automation artifact to be executed, for instance, source code or code comments. Second, we examined the~\textbf{input format}, which refers to how the automation artifacts collect the data (e.g., as \textbf{CSV} or \textbf{JSON} files). These two elements are important to understand how the automation artifacts work in practice and
also to support practitioners/researchers in choosing the appropriate tools (e.g., if one wants a dashboard with information about TD, it does not make sense to choose a tool that outputs a CSV). Moreover, these elements have not been previously studied in related work (see Section~\ref{sec:related_work}). Finally, they are important for further discussion about how the automation artifacts could be integrated (which will be presented in Section~\ref{sec:results_usage_automation artifacts}). 

\begin{figure}[ht]
     \centering
     \begin{subfigure}[b]{0.48\textwidth}
         \centering
         \includegraphics[width=\textwidth]{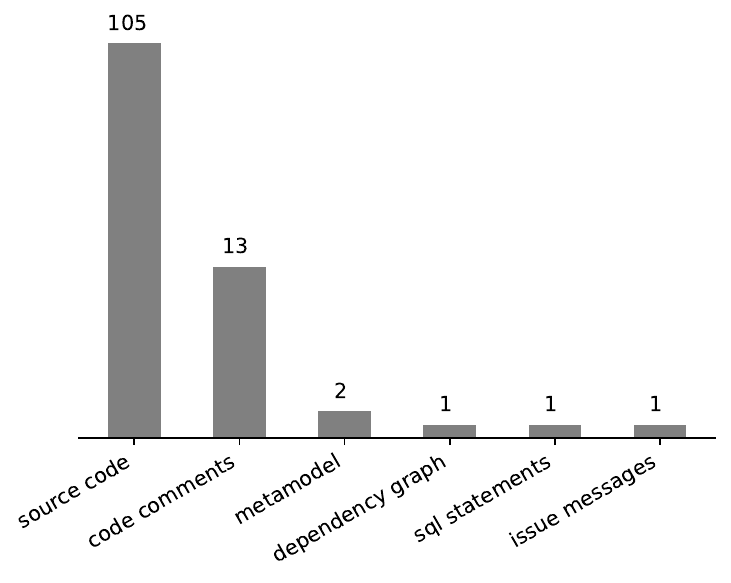}
         \caption{Inputs}
     \end{subfigure}
    \hspace{1em}
     \begin{subfigure}[b]{0.48\textwidth}
         \centering
         \includegraphics[width=\textwidth]{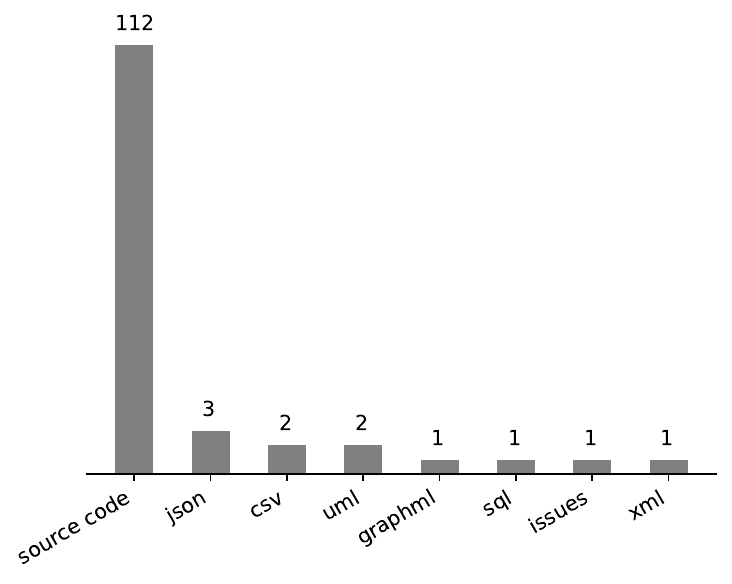}
         \caption{Input Formats}
     \end{subfigure}
     \hfill
        \caption{Inputs and formats that are used by the automation artifacts}
        \label{fig:automation artifact_input_type}
\end{figure}

Regarding the \textit{information}, we found six different types that are used by the automation artifacts, namely: \textit{source code, code comments, dependency graph, SQL statements, issue messages, and metamodels}, as presented in Figure~\ref{fig:automation artifact_input_type}a.

The majority of the automation artifacts (105/121) use \textbf{source code} as information to proceed with the automation. Specifically, automation artifacts that use this information focus on capturing data related to the quality of the code base, such as metrics and smells. This information is similar to findings of recent studies that present the stakeholders' focus on identifying TD considering the source code~\cite{Besker2018} (we will further discuss the TD types and activities in Section~\ref{sec:results_tdm_activities}). Source code can be used to support the management of all TD types (except for infrastructure and versioning, which are not supported by any tool). Specifically, 82/105 artifacts that use source code as input can support code TD. Similarly, 52/105 support design TD and 27/105 support architectural TD. The remaining types are supported in a much lower level.

Secondly, \textbf{code comments} are used by around 10\%~(13/121) of the automation artifacts. Code comments are used to deal with SATD, which are commonly documented through this type of data. Moreover, SATD is also detected using \textbf{issues and commit messages} by 1 automation artifact out of 121. In addition, 1 automation artifact out of 121 uses \textbf{dependencies graph} as input information to, support the management of architectural TD.To analyze database design TD, 1 automation artifact out of 121 also uses \textbf{SQL statements}. Finally, 2 automation artifacts out of 121 that employ model-driven development techniques, use \textbf{metamodels} to identify design TD in the models.

Regarding the \textit{format} used by the automation artifacts, we found eight different types, as presented in Figure~\ref{fig:automation artifact_input_type}b. \textbf{Source code files} are directly accessed by 112/121 automation artifacts. Specifically, 92/105 use the source code as information, and 13 extract the code comments from the source code. This usage can be done through accessing the projects locally (66/92) or remotely (8/92), integrated with platforms such as GitHub. Moreover, the source code can also be accessed directly through IDEs, by plugins (33/92). Two automation artifacts use pre-processed \textbf{CSV} files, three use \textbf{JSON} to perform the automation, related to the identification of SATD, while one uses GitHub issues (also for SATD). Finally, \textbf{XML} and \textbf{UML} files are used by two automation artifacts to perform the analysis of metamodels.

There are 25\footnote{The list with all types can be found in out replication package: \replicationpkgurl} types of output information, which are provided by the automation artifacts after its execution. Around 30\% (43/121) of the automation artifacts can calculate code metrics (e.g., cyclomatic complexity), which represent the most cited output information. Code smells (28/121) are also well supported. Besides, around 10\% (14/121) of the automation artifacts provide a list of SATD items, both considering the code comments and the issue messages.

Regarding the output formats, we found that most automation artifacts provide their results into textual format. The most common output format is text (e.g., txt), supported by 37/121 automation artifacts. Structured formats, such as  xml (28/121), csv (26/121), and json (25/121), are also supported by the automation artifacts. HTML is supported by 22/121 automation artifacts, which usually provide reports that can be accessed through browsers.

\vspace{6pt}
\noindent\fbox{%
    \parbox{\textwidth}{%
    \textbf{\textit{Finding 3:}} Source code is the most common input used by the automation artifacts. It is followed by code comments, although with a much lower level of attention and restricted to identifying SATD. Other types of input, such as metamodels, are also used. Regarding the outputs, code metrics is prevalent among output types, followed by code smells. SATD items are also the output of several automation artifacts.
    }%
}

\subsubsection{Evidence level of the automation artifacts}
\begin{figure}[ht]
\centering
    \includegraphics[width=\textwidth]{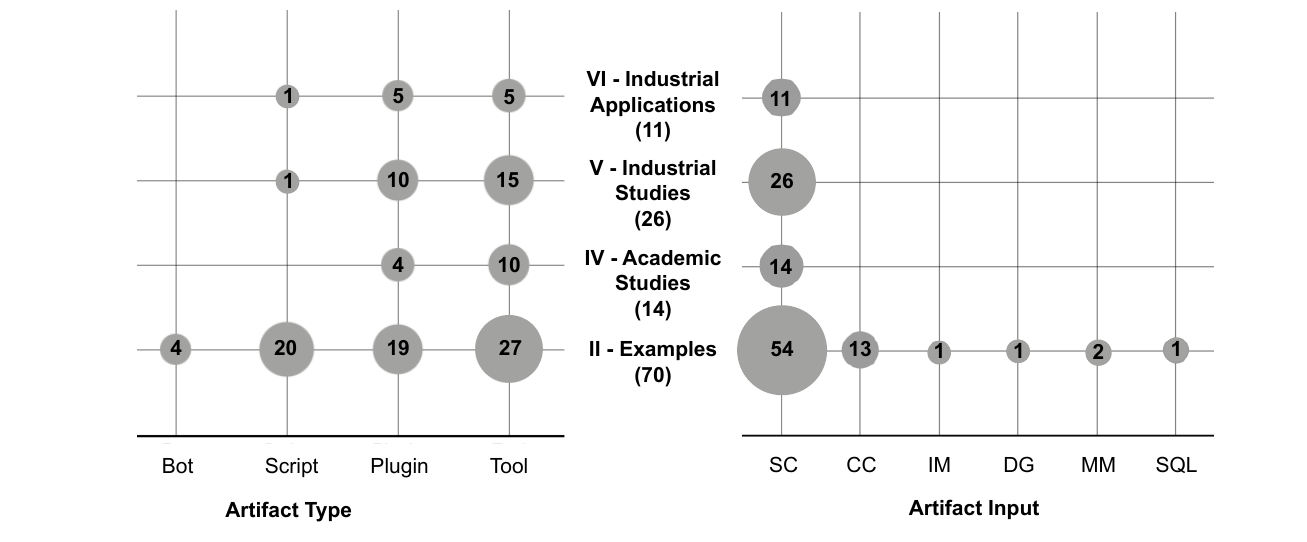}
    \caption{Evidence Level}
    \label{fig:automation artifact-evidence-level}
\end{figure}

In \textbf{RQ1.3}, we aimed at understanding the evidence level of the automation artifacts.
For this classification, we adopted the six levels of evidence proposed by Alves et al.~\cite{Alves2010_evidence_level}: (I) no evidence; (II) examples; (III) expert opinions; (IV) academic studies; (V) industrial studies; and (VI) industrial applications. To categorize the automation artifacts we took into account the evidence we found in the primary studies. Consequently, some automation artifacts that may be currently used in the industry, but have no published evidence, were not classified as ``Industrial Applications''. BetterCodeHub\footnote{\url{https://bettercodehub.com/}} (now known as Sigrid\footnote{\url{https://www.softwareimprovementgroup.com/}}), a static code analyzer, is one such example. Although there is information about industrial applications in the tool's website, 
we only found it as an example in the studies, so it was classified in Level II. 
While this strategy of classification could underestimate the level of maturity of a tool (as presented in the previous example), it ensures that our analysis considers the evidence we found in peer-reviewed studies, improving the reliability of this work.
Figure~\ref{fig:automation artifact-evidence-level} depicts the number of automation artifacts classified in each category.

From the information presented in Figure~\ref{fig:automation artifact-evidence-level}, automation artifacts with evidence Level II (\textbf{examples}) represent around 60\% (70/121) of the total. In this category, we considered automation artifacts that were used or just proposed  in a study, but not evaluated. 
Level IV (\textbf{academic studies}) includes 11\%~(14/121) of the automation artifacts and represents automation artifacts that were evaluated during academic studies i.e., besides their application in an academic study, their correctness and efficiency were analyzed. Furthermore, 20\% (26/121 automation artifacts) were used in \textbf{industrial studies}, referring to Level V of evidence. At this level, the automation artifacts are evaluated in an industrial context. Finally, we found evidence that 8\% (11/121) of the automation artifacts reached \textbf{industrial application} (i.e., Level VI of evidence). This level refers to automation artifacts that are used by the companies during software development.Finally, it is important to highlight that we only classified automation artifacts in the mentioned four different levels of evidence; none of the studied automation artifacts was classified into levels I (no evidence) or III (expert opinions). Every automation artifact was applied by the authors in at least one study. For Level III, we did not find any automation artifacts being evaluated by experts.

In Figure~\ref{fig:automation artifact-evidence-level}, we also presented the level of evidence of the automation artifacts, considering their types. To a great degree, the usage of tools is related to higher levels of evidence i.e., \textbf{Levels IV and V}. Since those levels are directly related to industry, 
these results can indicate that configuration and flexibility capabilities seem to be prioritized in industrial contexts. In the same direction, several plugins also reached higher levels of evidence (IV and V). Their main characteristic is the easy integration in the work environment (since they can be installed in IDEs, for example), which may imply that a smoother integration between the automation artifact and the development environment is prioritized in an industrial context. In the same vein, we found two scripts that reached higher levels of evidence (\textit{cbr insight} and \textit{piranha}). Those scripts were developed inside the companies and applied to a specific context; presumably, they are easy to use in this context, without requiring much configuration.

Finally, Figure~\ref{fig:automation artifact-evidence-level} also provides information about the input used by the automation artifacts (i.e., \textbf{S}ource \textbf{C}ode, \textbf{C}ode \textbf{C}omments, \textbf{I}ssue \textbf{M}essages, \textbf{D}ependency \textbf{G}raph, \textbf{M}eta\textbf{m}odels, SQL). Those that take source code as input reach higher levels of evidence (approx. 30\% of the artifacts, incl. levels V and VI). Some automation artifacts (18/121) in Level II use information other than source code (e.g., code comments and issue messages).

In RQ1, we review the main characteristics of automation artifacts that can support TDM. Specifically, we classify the artifacts into four categories (i.e., tools, scripts, plugins, and bots) and highlight their main characteristics, which is a contribution that was not identified in previous studies. Moreover, we investigate technical aspects, such as inputs and outputs, which was also not performed before and can support the analysis and selection of artifacts by practitioners. Finally, similarly to previous studies~\cite{Avgeriou2021-fi, Silva2022}, we classify the artifacts considering their level of evidence, helping in understanding how artifacts were used and/or evaluated (e.g. evaluated in academic studies) and in which context (e.g., used for industrial
applications.

\vspace{6pt}
\noindent\fbox{%
    \parbox{\textwidth}{%
    \textbf{\textit{Finding 4:}} The most common level of evidence (approx. 60\%, 70/121) is Level II (i.e., examples). Nonetheless, many automation artifacts are used or evaluated in an industrial context (approx. 30\%, 37/121). Finally, automation artifacts with higher levels of evidence tend to use source code as input.
    }%
}

\subsection{Automation of TDM Activities and TD Types}
\label{sec:results_tdm_activities}

In \textbf{RQ2}, we aimed at understanding how the identified automation artifacts have been applied to automate TDM. Specifically, we explored the TD types and TDM activities for which we found support. We emphasize that some automation artifacts can handle more than one type of TD. For example, SonarQube can deal with four types of TD (architectural, design, code, and test), so it is counted five times in the discussion of automation artifacts for TD types (one for each type). Similarly, some automation artifacts are able to deal with more than one activity. For instance, Codacy\footnote{\url{https://www.codacy.com/}} can both \textit{identify} and \textit{monitor} TD items related to code TD. In this case, during the discussion of TDMA, it is counted twice (once for identification and once for monitoring).  

\begin{figure}[ht]
    \centering
    \includegraphics[width=0.72\textwidth]{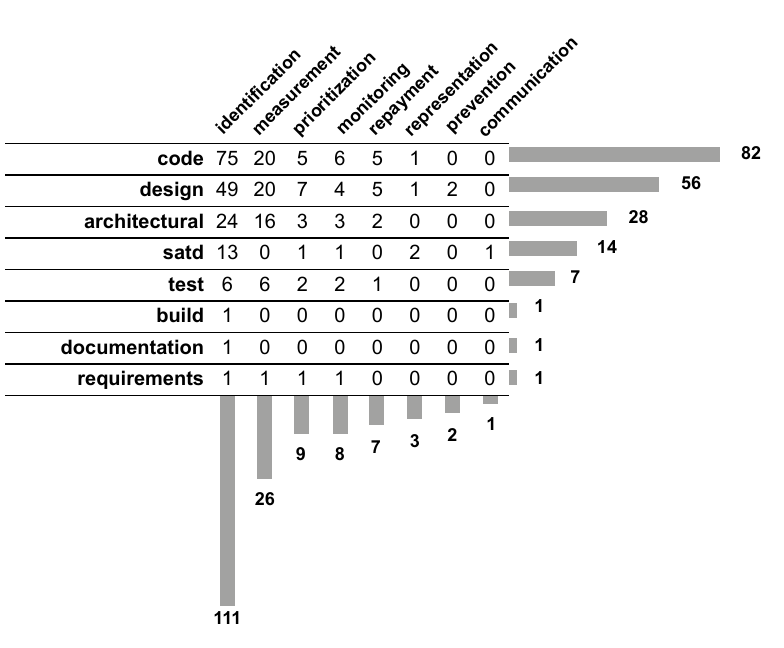}
    \caption{Number of automation artifacts that support each TDM activity for each one of the TD types}
    \label{fig:td_types_tdma}
\end{figure}

Figure~\ref{fig:td_types_tdma} summarizes the number of automation artifacts available for each TD Type and TDM Activity.
Several studies~\cite{Junior2022,Li2015,Alves2016} discuss the relation between the type of TD and the tasks to manage it, which highlights the need for aligning the automation process with the TD type. Thus, we will discuss how the automation artifacts can deal with the TDM activities, for each TD type. \textbf{Code TD} is the most supported type by automation artifacts, with 63\% (82/121) being able to handle at least one TDM activity for this type. The \textit{Identification} activity is the most automated, since 75/82 automation artifacts deal with the identification of Code TD. \textit{Measurement} is the second one, with 20/82. Other activities also received some attention but at a much lower level, such as \textit{Repayment}~(5/82) and \textit{Monitoring}~(8/82). We did not find any evidence related to the automation of \textit{Prevention} in our results, which may suggest a focus on dealing with the existing TD, and not necessarily avoiding new TD items. However, we identified several plugins that can be used alongside IDEs, providing real-time feedback, which could reduce the insertion of new TD items. However, these automation artifacts do not explicitly support prevention but we think they have potential in doing so if properly extended.

Considering \textbf{Design TD}, 45\% (58/121) of the automation artifacts were used in the automation of activities related to this type of TD. 49/58 were used to \textit{identify} design problems. Besides, 20/48 automation artifacts can \textit{measure} the amount of Design TD present in a software system. The \textit{Prioritization} activity is supported by 7/58 automation artifacts, which is the highest number of prioritization-related automation artifacts among all TD types. A potential reason may be related to 
the potential ripple effect of poor design decisions (i.e., they can cause several other problems in the software), which drives stakeholders to prioritize and repay design TD items~\cite{Fontana2016-ta}.
Besides, design TD is the only type where \textit{Prevention} activity is automated (2/58). Regarding \textit{repayment}, it is considered more challenging to automatically repay design TD (and architectural TD) when compared with other types (e.g., code TD), due to the complexity and extent of the repayment.  
Nonetheless, we found 5/58 automation artifacts that can automatically \textit{repay} design TD: 4/5 refer to removing code smells from the source code and 1/5 to removing design smells from metamodels. This last one, specifically, can act as an alternative to handling design TD in the early stages of software development.

An example of an
automation artifact for automated repayment of architectural TD is derec-gea~\footnote{\url{https://github.com/teomaik/DeRec-GEA}}, which relies on the identification of modularity issues (a specific type of architectural TD~\cite{Besker2018}); it can can suggest refactoring actions, but is still up to the stakeholders to accept it. In contrast to source code, we did not identify any tool capable of refactoring architectural TD without human supervision; presumably because architectural changes are complex and bear risks.

\textbf{Test TD} is supported by 7/121.
\textit{Identification} is still the most considered activity, with 6/7 automation artifacts for Test TD. The identification of the TD items related to Test TD is made by checking the test coverage of the software. Moreover, the automated \textit{Measurement} of Test TD is provided by 6/7 automation artifacts. 

\textbf{Requirements TD}, \textbf{Documentation TD}, and \textbf{Build TD} are supported by one automation artifact each. This may suggest that stakeholders tend to pay less attention to those types of debt, compared to the rest. Regarding \textbf{Requirements TD} and \textbf{Documentation TD}, 
another explanation could be that those types of TD cannot be identified using source code, which is widely used as input to deal with TD.

Although we do not classify SATD as a TD type like the others, several automation artifacts are proposed and used specifically to deal with this kind of TD. Specifically, 14/121 automation artifacts can detect the presence of SATD in code comments and issue messages. Among those automation artifacts, 13/14 can \textit{identify} SATD items, 2/13 can \textit{represent} them, and 1/13 can \textit{communicate} them.

Since many automation artifacts can perform the same activity, it is worth reporting that some empirical studies compared the artifacts in terms of accuracy. Specifically, 25/178 artifacts were compared in 15 studies. Most of these artifacts (20/25) are used for \textit{identification} of  code debt. Overall, SonarQube is the one that presents more functionalities and the highest number of rules~\cite{Lefever2021-ot}, while its accuracy is not that high~\cite{Fontana2016-mt}. In terms of accuracy, Lefever et al.~\cite{Lefever2021-ot} present a study that compares three artifacts (Structure 101, SonarQube, and DV8) for code smells detection. Their results show that the anti-patterns detected by DV8, especially Modularity Violation and Unstable interface,
provide support to the identification of debt-ridden files, which points to a high accuracy of this artifact in this context. 11/25 artifacts were compared for \textit{measuring} TD. Specifically, Avgeriou et al.~\cite{Avgeriou2021-fi} conducted a study that compared nine tools for TD measurement and reported that NDepend, CAST, DV8, and CodeMRI can calculate TD interest, and would be recommended if the main priority is estimating extra effort for repaying TD. On the other hand, other tools are better when the priorities lie elsewhere, such as security (CAST and SonarQube), maintainability at multiple levels (CAST, NDepend, and SQuORE), and architectural analysis (CAST, NDepend, and SonarGraph ).

\subsubsection{Artifacts Types per TD Type and TDMA}

\begin{figure}[h]
\begin{minipage}[b]{\textwidth}
    \centering
    \includegraphics[width=0.52\textwidth]{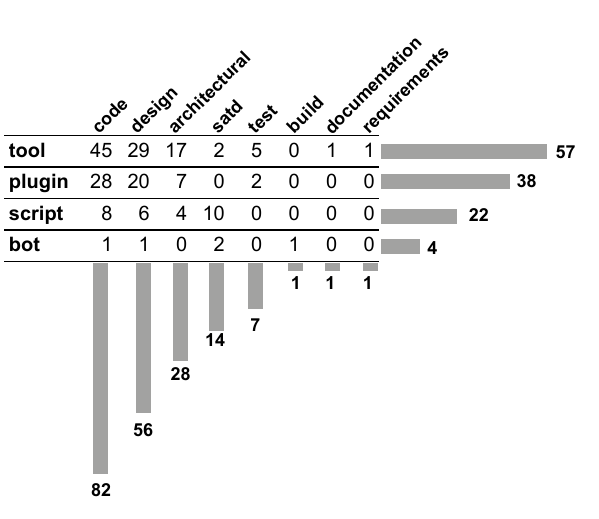}
    \caption{Types of automation artifacts per TD type}
    \label{fig:artifact-type-per-td-type}
\end{minipage}

\begin{minipage}[b]{\textwidth}
   \centering
    \includegraphics[width=0.52\textwidth]{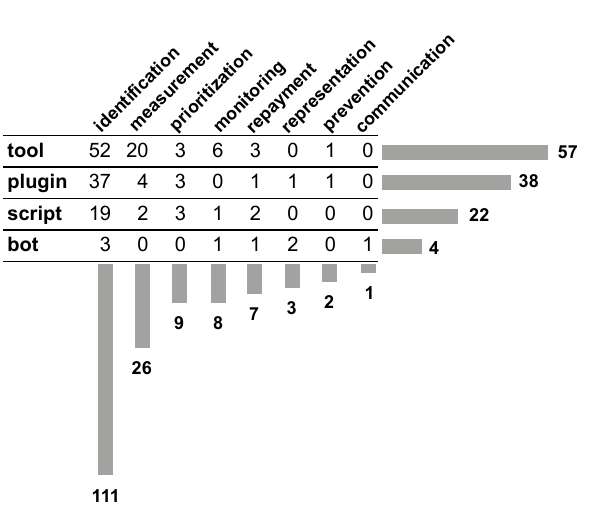}
    \caption{Types of automation artifacts per TDM activity}
    \label{fig:artifact-type-per-tdma}
\end{minipage}
\end{figure}

We also analyzed the relation between the type of artifacts and the TD types that are supported by them. Figure~\ref{fig:artifact-type-per-td-type} shows how many artifacts of each type support each TD type. We note that one same artifact can support multiple TD types, thus the sum of the numbers in each row does not match the total of automation artifacts.

Differently from other types of automation artifacts, \textbf{scripts} are mostly used for identifying SATD (10/22). This is possibly because these scripts are used to execute machine-learning models that analyzes natural language. Besides, scripts are also used for supporting the management of other types of TD, i.e., code (8/22), design (6/22), and architectural (4/22). For these types, the scripts usually query results from other artifacts to proceed analysis. 

Also, \textbf{bots} are the only type of artifact that supports build TD. Specifically, Breaking-bot~\cite{Ochoa2022} supports the management of libraries, and notifies the maintainers of OSS if some change in a certain library could break the system that uses this library as dependency.

Regarding TDM activities, Figure~\ref{fig:artifact-type-per-tdma} shows a similar distribution among the the types of artifacts. Most notably, identification is the most commonly supported activity, while \textbf{tools} are also prominently used to measurement.

\subsubsection{Relation among TD Activities and TD Types}
\label{sec:results_relation_tdma}

\begin{figure}
    \centering
    \includegraphics[width=\textwidth]{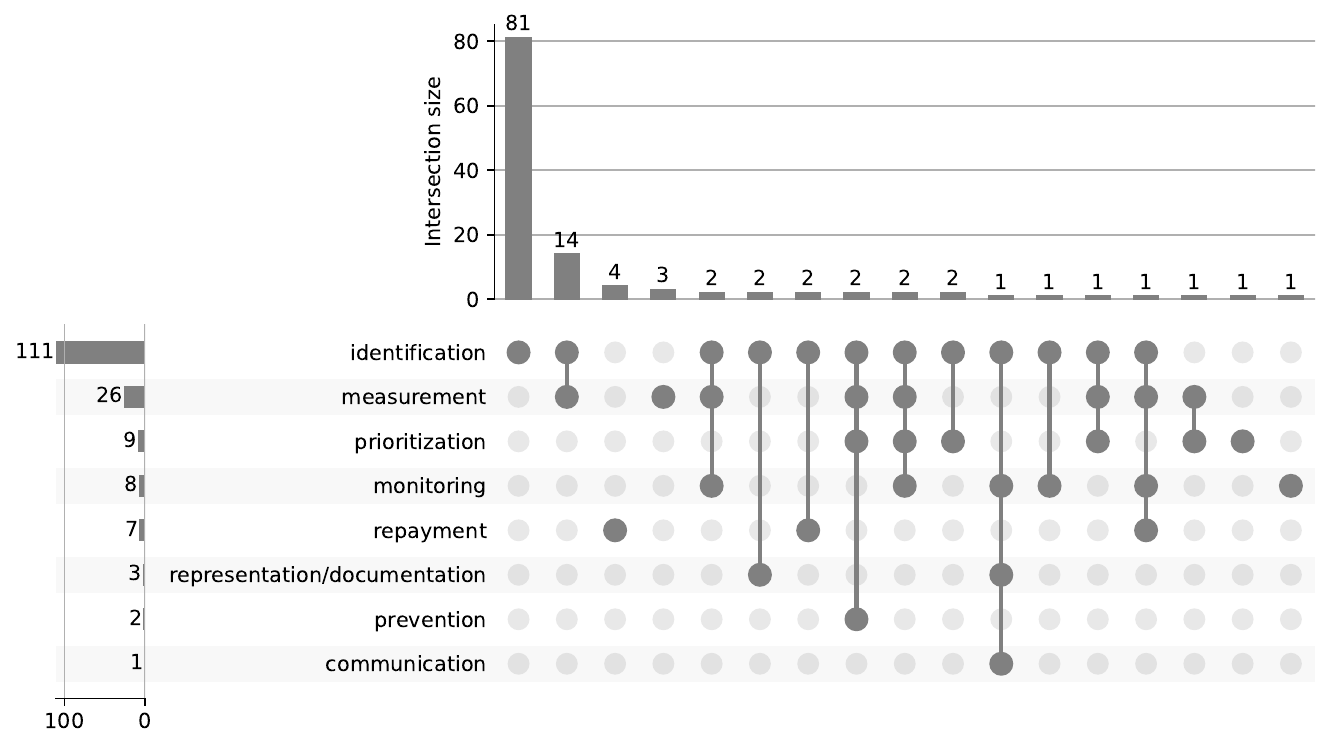}
    \caption{Number of automation artifacts that support one or more TDM activities.}
    \label{fig:relation_tdma}
\end{figure}

\begin{figure}
    \centering
    \includegraphics[width=0.8\textwidth]{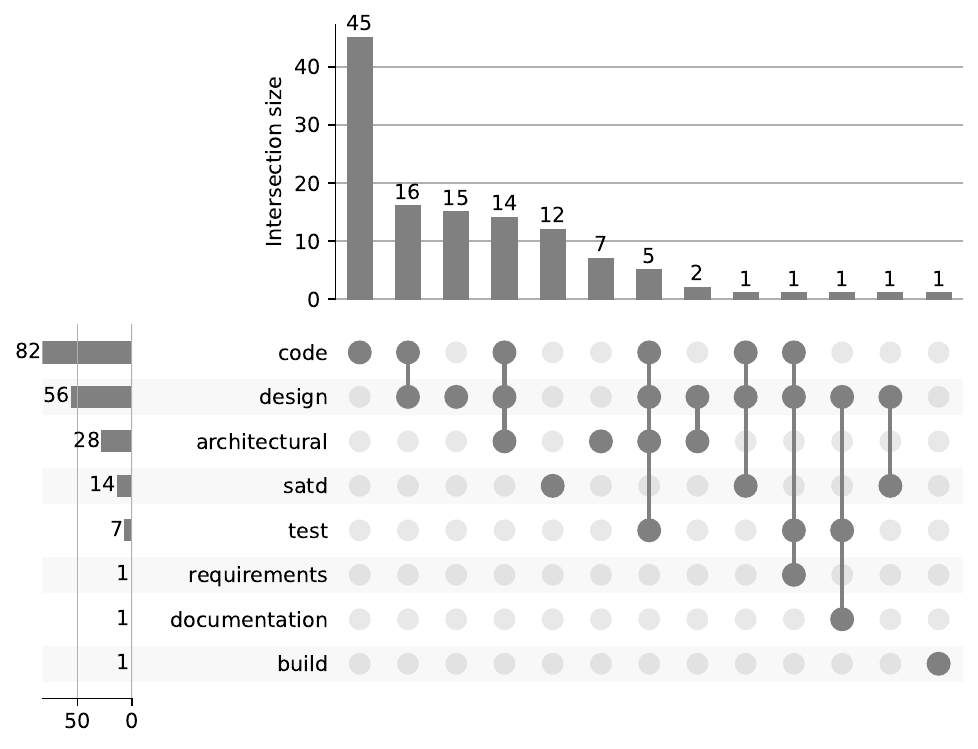}
    \caption{Number of automation artifacts that support one or more TD types.}
    \label{fig:relation_td_types}
\end{figure}

Although most automation artifacts (90/121) deal with a specific TDM activity, we found 31 automation artifacts that can perform two or more activities, as presented in Figure~\ref{fig:relation_tdma}. Among those automation artifacts, we identify just 1/31 that do not perform the \textit{identification} activity. This result indicates that the \textit{identification} of TD is present in almost all automation strategies, since it is necessary to list TD items before proceeding with the other TD activities. Moreover, the \textit{measurement} of TD is the most common activity that is combined with the \textit{identification} activity (14/31). Considering the TD types, 41/121 automation artifacts can support more than one type of TD, as shown in Figure~\ref{fig:relation_td_types}. Specifically, 16/41 automation artifacts can deal with code and design TD. Code, design, and architecture TD can also be managed by 14/41 automation artifacts, which also represents a strong relationship between these three types of TD.

In RQ2, we review the scope in which the automation artifacts can be used, specifically investigating the types of TD that are supported and the TDM activities that are automated. Previous studies~\cite{Khomyakov2019, Li2015, Silva2022} have already investigated this aspect. However, our study expands this classification by relating different TDM activities and TD types that are automated by a single artifact. This analysis sheds lights on the relations between TD types and TDM activities and can support the definition of TDM approaches.

\vspace{6pt}
\noindent\fbox{%
    \parbox{\textwidth}{%
    \textbf{\textit{Finding 5:}} The \textit{Identification} activity is the most supported activity of the TDM, while \textit{Measurement} also received significant attention. Moreover, the combination of those activities is present in 14/121 automation artifacts. Code, design, and architectural TD are the three types that received the most attention, while 14/121 automation artifacts can deal with all three of them. Finally, existing evidence regarding the performance and accuracy of artifacts is related to those artifacts that support \textit{Identification} and \textit{Measurement}.
    }%
}

\subsection{Usage of the automation artifacts}
\label{sec:results_usage_automation artifacts}

In \textbf{RQ3}, we analyze the evidence we found related to the usage of the automation artifacts regarding two main perspectives. The first one concerns the trigger (i.e., how the artifacts are executed), while the second regards the existing and possible integration between artifacts.

\subsubsection{Automation Artifacts' Triggers}

We classified the automation artifacts considering three different types of trigger:
(i) \textbf{human-based triggers}, i.e., requiring human intervention to be executed;
(ii) \textbf{automated triggers}, i.e., running automatically based on events that happen in the environment; and
(iii) \textbf{both}, i.e., automation artifacts that can be configured to run both with human-based and automated triggers. It is essential to analyze this aspect of automation artifacts to understand how they can be  executed in the development process, which enable as smoother integration of them in the existing development workflows that are present in industry.

\begin{figure}[ht]
    \centering
    \includegraphics[width=0.6\textwidth]{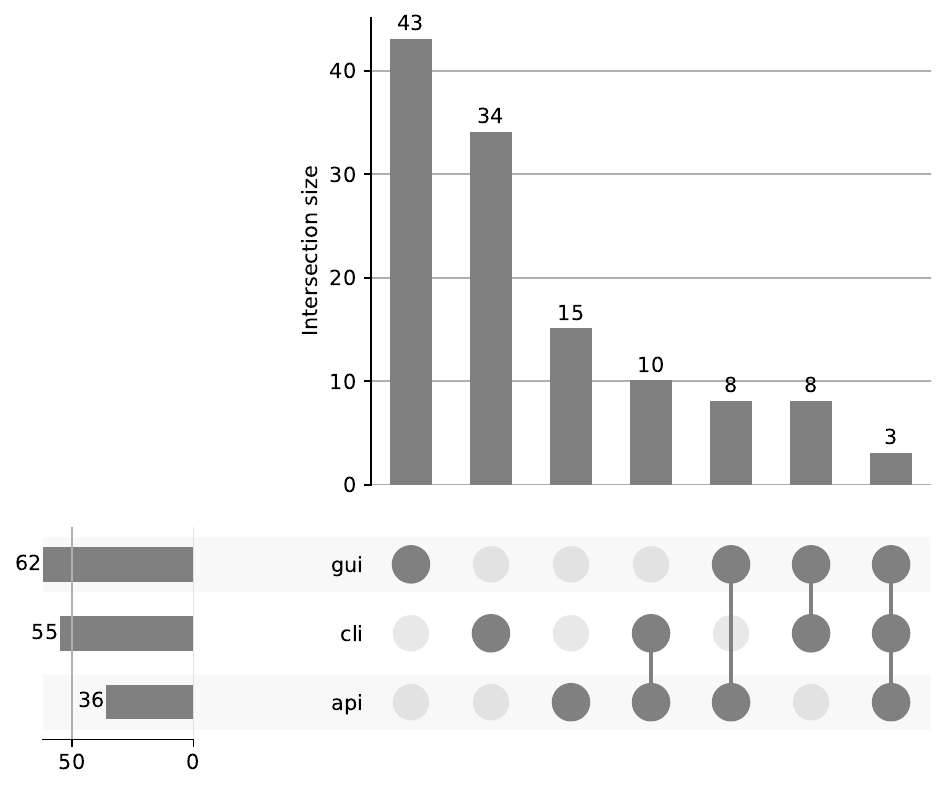}
    \caption{Type of interfaces used to execute the automation artifacts}
    \label{fig:rq3-types-of-interfaces}
\end{figure}

The classification was performed based on the interface(s) offered by the automation artifacts, which are presented in Figure~\ref{fig:rq3-types-of-interfaces}. Specifically, this figure presents the number of artifacts that provide each type of interface. Automation artifacts that provide only a graphical user interface (GUI) were categorized as human-triggered artifacts. We note that several plugins also contain human triggers since they are activated through a GUI element (e.g., a button added to an IDE via the plugin). The artifacts that can be executed only through an API were classified as automated triggers, as they require software integration. Automation artifacts classified under the category ``Both'' either: (a) offer a Command-Line Interface (CLI), i.e., can be executed manually via a terminal (e.g., Windows Powershell) or can be automated through Operating Systems process pipelines; or (b) present more than one type of interface (e.g., Kiuwan~\footnote{\url{https://www.kiuwan.com/}} provides an API and a GUI). 

\begin{figure}[ht]
    \centering
    \includegraphics[width=0.4\textwidth]{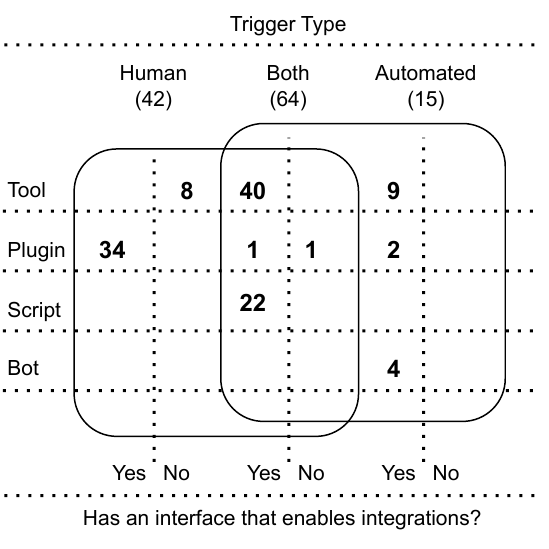}
    \caption{Type of the automation artifact x trigger}
    \label{fig:automation_artifact_type_trigger}
\end{figure}

Figure~\ref{fig:automation_artifact_type_trigger}  reports the triggers for the automation artifacts, classified by the type of artifacts (e.g., tool, plugin, etc) that were reported on RQ1. The figure also highlights the number of artifacts that present at least one interface that enables integrations (which will be further explored in RQ3.2). 
It shows that automation artifacts with support for both automated and human triggers are more common than the other types: approx. 53\% (63/121), compared to 35\% (43/121) for human triggers, and 12\% (15/121) for automated triggers.

To provide more insights on the usage of the automation artifact in the software development, we also analyzed how the types of triggers are related to the following set of activities: (a) requirements engineering; 
(b) architecture and design; 
(c) coding; 
(d) testing; and 
(e) deployment. 

Regarding \textbf{requirements engineering} and \textbf{deployment}, we did not identify any artifacts. Regarding \textbf{architecture and design}, we identified 58/121 automation artifacts support the management of architecture and design TD. For this activity, about half of the artifacts (27/58) present both human and automated triggers, and 25/58 provide only human-based triggers. Besides, 6/58 artifacts present only automated triggers.

\textbf{Coding} is the activity for which most of the automation artifacts can be used (99/121). This observation is not surprising since most artifacts take source code as input (see RQ1.2). For this set of artifacts, 54/99 provide both types of triggers, 31/99 provide only human-based triggers, and 14/99 provide only automated triggers.

Test TD, and by consequence \textbf{testing}, is supported by 7 artifacts. Three artifacts provide only human-based triggers, two provide only automated triggers, and two provide both.

\subsubsection{Integration between Automation Artifacts}

\begin{figure}[ht]
     \centering
     \begin{subfigure}[b]{0.32\textwidth}
         \centering
         \includegraphics[width=\textwidth]{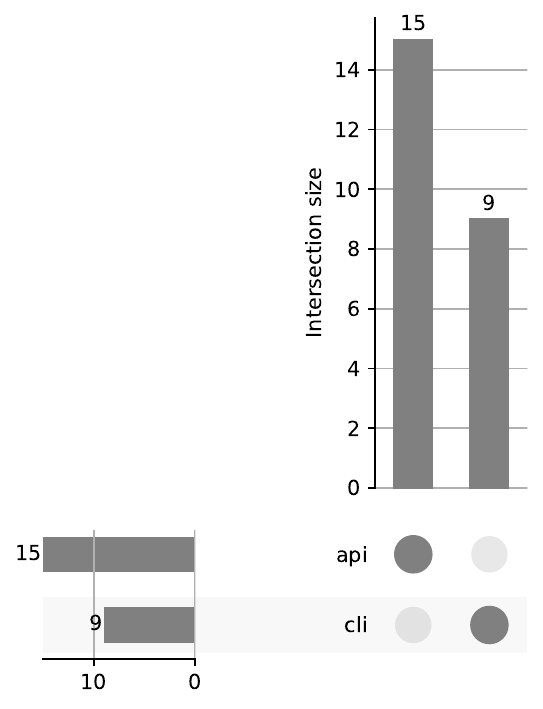}
         \caption{Number of artifacts and interfaces used in existing integrations}
         \label{fig:upset-existing-integration}
     \end{subfigure}
     \hspace{2em}
     \begin{subfigure}[b]{0.51\textwidth}
         \centering
         \includegraphics[width=\textwidth]{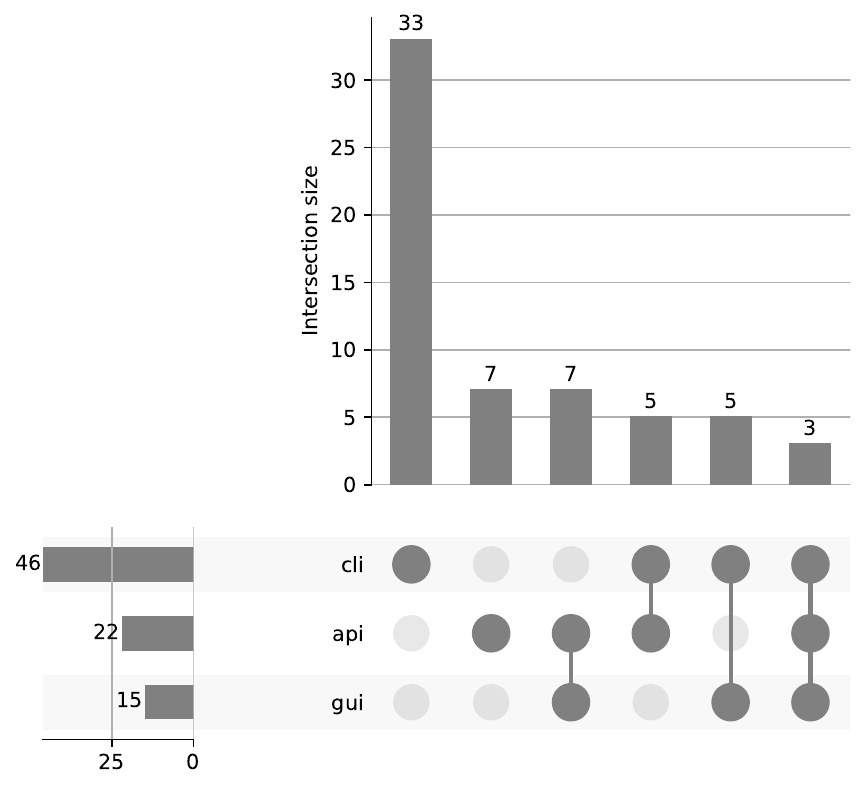}
         \caption{Number of artifacts that could be integrated and the available interfaces}
         \label{fig:upset-possible-integration}
     \end{subfigure}
     \hfill
        \caption{Number of automation artifacts per interface type, in (a) existing and (b) possible integrations}
        \label{fig:upsets-interfaces}
\end{figure}

The second perspective is related to the possibility of integrating different automation artifacts to improve the support for different activities and TD types (RQ3.2). For this purpose, we analyzed: a) the existing integrations between automation artifacts as suggested or performed in the primary studies; and b) possible integrations, which we inferred from the interface(s) provided by the  automation artifacts. Figure~\ref{fig:upsets-interfaces} presents the number of artifacts per type of interface. We note that automation artifacts may contain multiple interfaces from which only one was used in the corresponding primary study to integrate the artifact. In such cases, the used interfaces are accounted for in Figure~\ref{fig:upset-existing-integration} and the remaining ones in Figure~\ref{fig:upset-possible-integration}. 

Looking at the existing integrations, we found evidence of integrations for 24/121 automation artifacts, as presented in Figure~\ref{fig:upset-existing-integration}, from which 15/24 were made using API interfaces and 9/24 using CLI interfaces. The integrations were either with other automation artifacts (15/24) or other software (e.g., code hosting platforms such as GitHub (11/24) (which also includes bots, such as FixMe~\cite{Phaithoon_2021}), and IDEs or CI/CD Managers (8/24)). We note that some automation artifacts were integrated with more than one type of software: 8/24 artifacts were integrated with IDEs and other automation artifacts; 3/24 were integrated with automation artifacts and code hosting platforms; finally, 1/24 was integrated with code hosting platforms and CI/CD Managers. 

Among these existing integrations, we found that some of them support more than one TDM activity (13/14) or more than one TD Type (9/14). One notorious example is between \software{VisminerTD}\footnote{\url{https://visminer.github.io/}} and \software{Checkstyle}~\cite{Mendes_2019}, which enables the identification and measurement of TD (performed by Checkstyle), and visualization of the TD (through VisminerTD). In this case, the integration is done via API, in which VisminerTD uses Checkstyle's API to query the detected TD items. Such examples mean that researchers already started to look at the possibility of improving the automation process by connecting two or more automation artifacts. 
However, we note that several integration scripts had to be developed to connect the automation artifacts, which can be time-consuming and a trade-off that should be considered.

Some of the existing integrations concern integrating automation artifacts with other types of software (e.g., CI/CD Managers), thus enabling the execution of TDM activities in parallel with software development tasks. The integration between Jenkins~\footnote{\url{https://www.jenkins.io/}} (CI/CD manager) and CodeSonar~\footnote{\url{https://www.grammatech.com/codesonar-cc}}, for instance, enables the identification of TD while the software is deployed.

Similarly to RQ3.1, we also analyzed which development activities are supported by integrated automation artifacts. Specifically, 18/24 artifacts were integrated to support coding activity and 6/24 to support architecture/design activity. As we mentioned before, the integrations are usually employed to perform more than one TDM activity or to handle more than a TD type. For instance, Sas et al~\cite{Sas_2019} report that Arcan was integrated with Astracker to support the identification and monitoring of architecture TD. 

As for the possible integrations, i.e., those that we inferred from the interfaces (Figure~\ref{fig:upset-possible-integration}), we analyzed whether automation artifacts facilitate integration by providing a CLI or API. We found that 60/121 artifacts could be integrated, e.g., with IDEs, other automation artifacts, code hosting platforms, or CI Managers. 
Among these artifacts
, 46/60 present a CLI interface and 22/60 present an API interface. Besides, 15/60 artifacts present both an interface that enables integration and also a GUI interface. We also note in Figure~\ref{fig:upset-possible-integration} that 5/78 automation artifacts present both API and CLI interfaces that were not used in integrations in primary studies yet.

In RQ3 we explored technical aspects of automation artifacts, such as their triggers and possible integration. The main contribution of this RQ is to provide information on how those artifacts work and how they could be integrated in the software development process (e.g., integrated with code host platforms). Moreover, we highlight the high level of human interaction that is needed to execute the artifacts, which points out to the preference of tool support for TDM instead of higher levels automation.

\vspace{6pt}
\noindent\fbox{%
    \parbox{\textwidth}{%
    \textbf{\textit{Finding 6:}} 
    The execution of the majority of automation artifacts, can be triggered by 
 both human intervention or automatic strategies (e.g., using scripts) (approx. 53\%, 64/121). Besides, around 20\% (24/121) of the automation artifacts have been used in integrations, while 78/121 could be potentially integrated considering the interfaces they provide.
    }%
}

\section{Discussion}
\label{sec:discussion}

Our results show that researchers have been actively and intensively reporting the usage of automation artifacts to support TDM.
We further perform a synthesis of our results, comprised of a conceptual model for TDM automation, two examples of the model's usage, and a set of four challenges that should be addressed to improve TDM automation.

To define the conceptual model, initially we summarized the main results of each RQ. The results were then analyzed and the concepts and relations were defined based on constant comparison~\cite{Corbin2008, Stol2016}. 
For instance, RQ3 explores the usage of automation artifacts and shows that, usually, they are executed under certain rules (e.g., \software{TODO Bot}~\cite{Zampetti_2021} is executed after each commit).
This lead us to define the concept \modelconcept{TDM Automation Rule}.

Finally, it is relevant to highlight that the  model encompasses both the current state of TDM automation, but also our 
proposals on how to improve it, based on the limitations we identified in the current solutions for TDM automation. One example is the proposal of a central management for multiple automation artifacts with an \modelconcept{Orchestrator}.

\subsection{A Conceptual Model for TDM Automation}
\label{sec:theory}

\begin{figure}[ht]
    \centering
    \includegraphics[width=\textwidth]{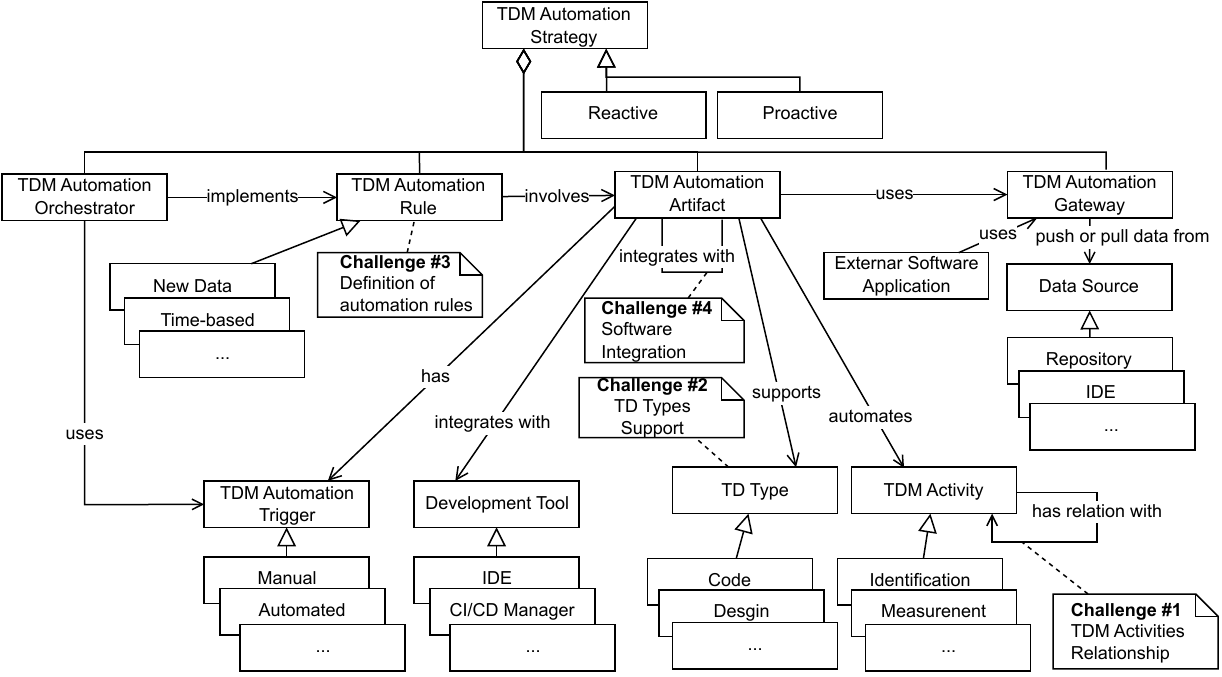}
    \caption{Conceptual Model of TDM Automation}
    \label{fig:theory}
\end{figure}

Similar to other conceptual models in the software engineering field (such as the one presented in Junior et al~\cite{Junior2022} for TD and TDM), a conceptual model for TDM automation can help to abstract and explain this domain. This model can aid practitioners' decision-making in specifying the TDM activities they would like to automate into a strategy and integrate those activities into an existing development workflow. In addition, since TDM cross-cuts different areas of a company (e.g., software development, project management, finance), the development of a strategy based on the model can improve communication among stakeholders. Figure~\ref{fig:theory} depicts the TDM automation conceptual model and 

Table~\ref{tab:theory_description} shows a summary of all concepts, including how they were identified and from which RQ.

The model is centered around the concept of a \modelconcept{TDM Automation Strategy}, which expresses how an organization or software development team addresses TDM automation. Such a strategy is composed of four principal concepts. The first concept is the \modelconcept{TDM Automation Artifact}, which refers to software that can automate one or more \modelconcept{TDM Activities}, and support one or more \modelconcept{TD Types}. More than one \modelconcept{TDM Automation Artifacts} can be used in a strategy, i.e., two or more of these artifacts can be integrated to provide more features for the strategy; this is expressed with the relation \textit{``integrates with.''} \modelconcept{TDM Automation Artifacts} can also be integrated with one or more \modelconcept{Development Tool}, such as IDEs, CI/CD Managers, and code hosting platforms (such as Github). Every automation artifact is initiated by an event (\modelconcept{TDM Automation Trigger}). A trigger can be manual (i.e., executed by human) or automated (e.g., software invocation), as presented in Section~\ref{sec:results_usage_automation artifacts}. We note that an automated trigger can also refer to the execution of one automation artifact by another, when two or more automation artifacts are integrated.

The second concept is the \modelconcept{TDM Automation Rule}, which defines when automation should take place (e.g., after a commit, or based on a threshold such as time or amount of TD items) and which \modelconcept{TDM Automation Artifact} should be executed.

The third main concept is the \modelconcept{TDM Automation Orchestrator}, a software that manages the strategy. Note that the orchestrator does not automate any \modelconcept{TDM Activity}; this is exclusively done by the \modelconcept{TDM Automation Artifacts}. Instead, it implements several \modelconcept{TDM Automation Rules} for the automation and uses the \modelconcept{TDM Automation Triggers} of the artifacts to execute those artifacts. 

The fourth concept is the \modelconcept{TDM Automation Gateway}, which serves as an interface to collect data to be used by the automation artifacts present in the strategy (e.g., collecting source code from repositories). The gateway can be used by a \modelconcept{TDM Automation Artifact} to query \modelconcept{Data Sources}, such as repositories, issue trackers, and IDEs (e.g., through plugins). The gateway can also be used by an \modelconcept{External Software Application}, which can query the gateway to retrieve data related to the automation results (e.g., a script that retrieves data or reports provided by the strategy). Also, during its execution, a \modelconcept{TDM Automation Artifact} may use the gateway to push results outside the strategy (e.g., creating pull requests).

Finally, a strategy can be implemented in two ways depending on how the orchestrator is used. In a \emph{reactive strategy}, the orchestrator will trigger the strategy based on a change in one or more data sources. In a \emph{proactive strategy}, the orchestrator triggers a strategy independently of the changes that occur to the data sources (e.g., the rules for automation are based on time or set manually by a developer).

\begin{xltabular}{\textwidth}{p{2cm}p{3cm}p{6cm}p{1cm}}
\caption{Concepts related to TDM Automation.}
\label{tab:theory_description}

\\ \hline \multicolumn{1}{c}{\textbf{concept}} & \multicolumn{1}{c}{\textbf{description}} & \multicolumn{1}{c}{\textbf{explanation}} & \multicolumn{1}{c}{\textbf{rq}} \\ \hline 
\endfirsthead

\multicolumn{3}{c}%
{\tablename\ \thetable{} (continued)} \\
\hline \multicolumn{1}{c}{\textbf{concept}} & \multicolumn{1}{c}{\textbf{description}} & \multicolumn{1}{c}{\textbf{explanation}} & \multicolumn{1}{c}{\textbf{rq}}\\ \hline
\endhead

\endlastfoot

TDM Automation Strategy 
    & Refers to how TDM automation should be addressed in software development. 
    & Overall, the results of this SMS suggest that automation artifacts have been applied individually, focusing on automating a specific TDM activity for a TD Type. However, the automation of TDM involves other aspects, such as the choice of automation artifacts, their integration, and the management of the automated solution. Thus, a complete strategy must be planned for TDM automation.
    & RQ1\newline RQ2\newline RQ3\\ 

TDM Automation Orchestrator
    & A software responsible for managing the TDM Automation Artifacts in the strategy.
    & Section~\ref{sec:results_usage_automation artifacts} discusses the integration between software. An example of this integration is \software{VisminnerTD}~\cite{Mendes_2019} and \software{SDK4ED}~\cite{Lamprakos_2022}. Although several automation artifacts are connected inside those platforms, they lack some kind of automated orchestration that implements rules without requiring developer intervention.
    & RQ3 \\

TDM Automation Rule 
    & Defines when the automation happens and which TDM Automation Software is used.
    & Section \ref{sec:results_usage_automation artifacts} summarizes how the automation artifacts are used in software development. It shows that some artifacts are executed manually. Since the main advantage of an automated strategy is to reduce the developers' workload, the definition of rules is crucial to specify how the automation artifacts should behave in a certain strategy; defining when an automation artifact must be executed (e.g., once a week, after each commit, etc) is an example of a rule.
    & RQ2, RQ3 \\

TDM Automation Artifact 
    & Software application that automates one or more TDM Activities.
    & To automate the TDM, one or more pieces of software must be used. Section~\ref{sec:results_sw_artifcats} summarizes the list of identified automation artifacts and Section~\ref{sec:results_usage_automation artifacts} discuses potential integrations between them.
    & RQ1\newline RQ3\\

TDM Automation Gateway
    & Responsible for pulling and pushing data in and out of the strategy.
    & As presented in Section~\ref{sec:results_sw_artifcats}, all automation artifacts need to interact with a data source to perform a TDM Activity. A gateway is then an adequate abstraction to isolate the strategy from the data sources.
    & RQ1 \\

Data Source
    & Refers to sources from which data is collected to be used within  the strategy.
    & Each automation artifact needs data to automate a TDM activity, as presented in Section \ref{fig:automation artifact_input_type}. Analyzing the data (and their format) is relevant in the strategy, for instance to evaluate possible data conversions that could be necessary (e.g., some artifacts can deal with code comments, which need to be extracted from source code collected from repositories).
    & RQ1 \\

External Software Application
    & Refers to software applications that are not within the strategy but can interact with it.
    & External software applications (e.g., scripts) can interact with the gateway to collect data from the strategy. For instance, scripts can be used to request reports about the automation process.
    & RQ1 \\

TDM Automation Trigger
    & It consists of the event and/or action that executes an automation artifact.
    & In Section \ref{sec:results_usage_automation artifacts}, we describe two types of triggers (manual and automated). It is important to analyze automation artifacts triggers to understand how they can be
    used to execute the artifacts in the strategy.
    & RQ3\\

Development Tool
    & Consist in tools that are used during software development, but are not used to automate TDM activities (e.g., IDE).
    & Section \ref{sec:results_usage_automation artifacts} explores the integrations in which automation artifacts are involved. As reported, the artifacts can be integrated with development tools, such as IDEs, CI/CD managers and code hosting platforms. Thus, analyzing such integrations during a strategy's design is relevant to facilitate the strategy`s integration into the software development (e.g., automation artifacts that can be integrated with IDEs could collect data directly from it).
    & RQ3 \\

TDM Activities
    & The activities that are automated in the strategy.
    & As presented in Section \ref{sec:results_tdm_activities}, the automation artifacts automate one or more TDM activities.
    & RQ2\\

TD Types
    & The types of TD that are supported by the strategy.
    & Similarly to TDM Activities, Section~\ref{sec:results_tdm_activities} shows that automation artifacts also support one or more types of TD.
    & RQ2 \\ \hline
\end{xltabular}

\subsection{Example of Usage}
\label{sec:pipeline}

\begin{figure}[h]
     \centering
     \includegraphics[width=\textwidth]{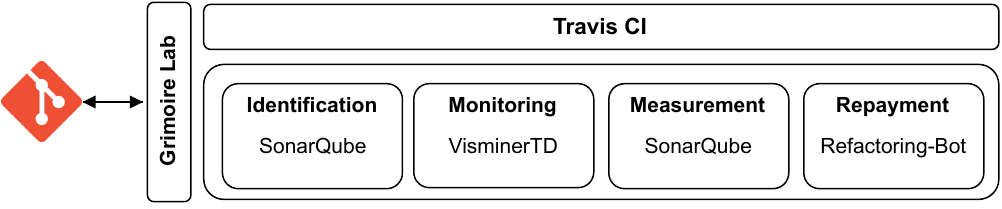}
    \caption{Example of TDM Automation Strategy}
    \label{fig:pipeline}
\end{figure}

To exemplify the usage of the conceptual model, we define an automation strategy, encompassing four of the discussed TDM activities. A conceptualization of the strategy is presented in Figure~\ref{fig:pipeline}.
As we showed in Section~\ref{sec:results_tdm_activities}, the type of TD drives the tasks that are performed to manage TD items. In this example we illustrate the management of Code TD.

In this example, we consider the TDM activities of  identification, monitoring, measurement, and repayment. In Figure~\ref{fig:pipeline}, we provide examples of software that could be used per activity. \software{SonarQube} is used for the \textbf{Identification} of TD, and this information is fed into \software{VisminnerTD}~\cite{Mendes_2019} for \textbf{Monitoring} TD through a dashboard. Subsequently,  \software{SonarQube} is applied for \textbf{Measurement}, i.e., to estimate the cost of paying back the TD. Finally, \software{RefactoringBot} is used in the strategy to pay back the TD Items identified by \software{SonarQube}. It is important to highlight that the data and interfaces provided by the automation artifacts would make possible their integration, mainly through accessing their interfaces (API and/or CLI), e.g., by using scripts that query data. However, this integration was not evaluated in practice.

In our conceptualization, \software{Travis CI}\footnote{\url{https://www.travis-ci.com/}} is the orchestrator and manages the automation process considering several rules. Moreover, \software{Grimoire Lab}\footnote{\url{https://chaoss.github.io/grimoirelab/}} acts as the gateway in this example, and collects data from repositories (e.g., \software{Github}), for use with automation artifacts that use  source code as input (e.g., performing static code analysis). 

There are two possible configurations for this strategy. In the first one, \software{Travis CI} manages the data collection, and upon some change in the data sources, triggers \software{Grimoire Lab} considering the rules that were defined in the strategy. This configuration enables a reactive implementation of the strategy, since the orchestrator can continuously check for changes in the data sources. 

In a second configuration, a proactive strategy is possible if data collection and orchestration are detached from each other. This means that the \software{Grimoire Lab} collects data continuously, making the data available through an API that could be consumed by the automation artifacts. The rules implemented in \software{Travis CI}, in this setting, do not depend on changes on the data sources; instead they could be periodically changed or manually by a user.

Looking into the implementation of the strategy, different vendors could provide parts of the solution. For instance, a vendor could offer a complete solution, including the data collection, orchestration, and TDM automation artifacts. Each part could also be developed independently and subsequently integrated (as the example in Figure~\ref{fig:pipeline} shows). Moreover, this flexibility can improve the reuse of tools, reducing the effort in developing the automation strategy.

Although the data provided by the selected automation artifacts, their interfaces and features theoretically enable the development of this strategy, this integration still presents some limitations, such as the support for just one specific type of TD. The limitations allude to four main challenges, which are discussed in the next section.

\subsection{Main Open Challenges}
\label{sec:challenges}

The conceptual model also highlights four main challenges that should be addressed in order to improve the level of TDM automation. \challenge{Challenge \#1} is ``\textit{How to automate various TDM Activities on the same strategy?}'', and mainly refers to the relation between TDM activities. As we showed in Section~\ref{sec:results_tdm_activities}, some automation artifacts can deal with more than one activity, which could provide an initial understanding on those relations. However, we still lack details on what the interdependence between the TDM Activities entails, when they are supported by one or more automation artifacts within a strategy. For instance, in software systems without documented TD, it would be impossible to automatically measure TD without identifying it before. Furthermore, depending on the approach to manage the TD items, after being identified they could be measured (e.g., if the primary goal of the approach is to obtain an overview on the TD accumulation) or repaid (e.g., simple code debt, such as lack of variable nomenclature names could be corrected without measurement) . This challenge directly affects the strategy's implementation since the order of activities must be mapped into the data exchanged between automation artifacts. Moreover, the decision of which activities should be automated in the strategy also drives the selection of automation artifacts.

In the same direction, \challenge{challenge \#2} is ``\textit{How to automate various TD Types on the same strategy?}'', and regards the handling of different types of TD within the same strategy. Defining which kind of TD will be managed is essential to determine the automation artifacts used. However, development teams often need to deal with multiple TD types since they can co-occur in software systems~\cite{Tan_2020b}. Some automation artifacts can deal with more than one type of TD, but the different types of TD are often investigated separately (as presented in Section~\ref{sec:results_tdm_activities}). In this sense, there is still a demand for approaches that exploit the best configuration for a strategy in order to combine more than one TD types. For instance, it could be possible to define several parallel strategies with different automation artifacts (to deal with different types of TD). However, this could increase the time and effort necessary to implement the strategy (see challenge \#4), which is a trade-off that should be considered and investigated.
Alternatively, automation artifacts able to deal with various types of TD could be chosen for a strategy. Besides, 
automation artifact orchestration could help in integrating artifacts and supporting more TD types. Thus, more attention in this direction is needed.

Currently, practitioners make several decisions related to managing different TD items, including which TD items should be repaid, which can stay in the system, and the trade-offs they consider to refactor the system. Thus, the role of the \modelconcept{Orchestrator} is played by a human in current TDM practices. However, in a automated TDM strategy (which aims at reducing human intervention) the \modelconcept{Rules} should be defined in advance and implemented within the \modelconcept{Orchestrator}. In this context, \challenge{challenge \#3} is ``\textit{What are the rules to automate TDM?}.'' 
Despite some research concerning prioritization strategies~\cite{Lenarduzzi2021}, we could not find application to an automated environment. Thus, it is necessary to understand to what extent the current techniques are applicable and, if necessary, propose new (reusable) rules based on specific metrics (e.g., coupling of the system) to specify how to proceed with certain TD activities and when (e.g., decide which TD items should be repaid). Moreover, the set of rules to perform a TDM activity should be clear and quantifiable since the orchestrator should be able to apply them without human intervention.

Finally, \challenge{challenge \#4} is ``\textit{How can we integrate TDM Automation Artifacts?}'', and is related to several technical issues when trying to integrate different automation artifacts in a strategy. 
The main technical problem is interoperability\footnote{By \textit{interoperability}, we mean the capacity of exchanging data between two or more automation artifacts.} between automation artifacts. As we presented in Sections~\ref{sec:results_sw_artifcats} and~\ref{sec:results_usage_automation artifacts}, available automation artifacts use different forms of input and output. When we look at the output provided by the automation artifacts, most of them offer common machine-readable formats (e.g., CSV), which theoretically makes the integration easy. However, the main issues are related to the input of the automation artifacts. In general, the automation artifacts use source code files as input, and cannot read files generated by other automation artifacts. In our example, Refactoring-Bot is used based on results generated by SonarQube. However, integration may not be possible if practitioners opted for another combination of artifacts instead of Refactoring-bot and SonarQube. To tackle this type of integration issue in the automation strategies, we need to: (i) develop new tools which can read data from several formats (e.g., JSON); or (ii) develop integration scripts that can translate results. 

Challenge \#4 also covers some execution problems of the automation artifacts. As we stated in Section~\ref{sec:results_usage_automation artifacts}, while most of the automation artifacts could be executed using both automated and human triggers, just few examples presented artifacts being executed by automated triggers. Thus, human intervention is still required in most cases, which hinders automation strategies that would aim at reducing human interaction as much as possible. Hence, more investigation is needed on how to improve the autonomy of the automation artifacts.. 

\subsection{Futures Perspectives of TDM Automation}
\label{sec:discussion_future_perpectives}

In this section we try to shed light on future research directions regarding the automation of TDM, based on the four challenges described in the previous section. 

To address \challenge{challenge \#1}, we argue that it is necessary to
establish the relations between TDM activities. To this end, we suggest two starting points. First, the results we presented in Section~\ref{sec:results_relation_tdma} show that some automation artifacts can perform more than one TDM activity, indicating that some TDM activities are more prone to be automated together. For instance, we found that all artifacts that \textit{document} TD items, also \textit{identify} them. While the opposite is not true, we can infer a certain dependency of the \textbf{documentation} activity on the \textbf{identification} one. Second, the concept of TD lifecycle~\cite{Zabardast_2022, Tan_2022b} refers to the steps that are followed to deal with TD items, and encompasses some TDM activities, such as identification and repayment.
It establishes the order that some TDM activities are executed, which can be understood as a type of relation among them, providing initial evidence on their inter-dependencies.

Regarding \challenge{challenge \#2}, different TD types should be supported to increase the level of automation of TDM strategies. 
A deeper exploration of the co-occurrence of TD types could provide an overview of what TD combinations should be supported, and could drive the choice of automation artifacts in a strategy. For example, if design and code TD usually co-occur, it would be pertinent to develop a single strategy that deals with both types. 
In an initial exploration of this topic, Tan et al.~\cite{Tan_2020b} studied how five different types of TD evolve in open-source Python projects. However, more evidence is required considering more TD types, other programming languages (e.g., C++), and in industrial settings.
Finally, it is necessary to explore the different strategies' configurations to deal with more than one TD type, through: a) exploring the usage of artifacts that support more than one TD type and b) exploring the integration of artifacts that support different types of TD.

To deal with \challenge{challenge \#3}, more research is needed to understand the decision-making process behind the automation of TDM activities (currently performed manually), i.e., the rules that software practitioners use to perform automated TDM activities. Specifically, we need to understand better the rationale of practitioners when they make those decisions, as well as the required automation triggers. For example, we need a better understanding of when a refactoring should happen (based on the accumulation of TD, the impact of the TD item, etc.). Moreover, we know very little about the practitioners' take on the negative aspects of automation. Understanding this, would help us to determine when or which decisions or tasks should \textit{not} be automated.

Finally, the technical issues related to \challenge{challenge \#4} should also be investigated. More effort in updating tools to follow standards could be an excellent first step in this direction. For instance, SARIF-based~\footnote{\url{https://sarifweb.azurewebsites.net/}} tools started to be used in the context of TDM, but this standard's usage is relatively low~\cite{Anderson2019-iv}. Besides, the Object Management Group (manager of UML) proposed a standard to automatically identify code debt~\cite{OMG2021}, which should also be investigated as a way to make it possible to integrate different tools. Moreover, we reported that some automation artifacts could be integrated into the CI/CD pipelines to help TD management (e.g., SonarQube). However, we did not find any evidence of this usage in practice.
Thus, it is also important to investigate why practitioners are not taking advantage of integration with CI/CD and what they would require to do so.

\subsection{Implications for Researchers}
\label{sec:discussion_implications_researchers}

First, a \textbf{deeper understanding of the automation process from the point of view of practitioners} must be provided. The 
conceptual model we proposed is based on information collected from automation artifacts identified through our SMS. However, most studies present researchers' viewpoints on the automation process. Therefore, to improve the body of knowledge and usability of the automation artifacts, it is crucial also to survey practitioners and understand their aims for the automation of TDM. 

Furthermore, we found some evidence related to the integration of automation artifacts, which enable supporting multiple TD types and TDM activities. However, not all automation artifacts that could potentially be integrated have actually been integrated in research approaches. Therefore, we argue that \textbf{it is essential to investigate new integrations between automation artifacts} in order to exploit all meaningful integration options and assess their feasibility within a TDM automation strategy.

Moreover, evidence regarding \textbf{the comparison between the evolution of TD in automated and non-automated settings} is still necessary.
It is important to investigate those scenarios to better understand in which of them the automation is more usefull. For instance, would be important to investigate if the time to repay TD items is shortened when automation artifacts are used.

From another perspective, we see the usage of SATD as a potential driver to improve TDM automation. This is because SATD items are already documented, and can help locating real problems in the software system. However, two directions need to be investigated to make the usage of SATD even more helpful. First, the identification of SATD can be time-consuming and the accuracy of current classifiers is not particularly high (e.g.,  61\% reported by Li et al.~\cite{Li2023}). Thus, \textbf{more research in improving the accuracy and efficiency of classifiers is needed}.
Second, when considering the orchestration of different automation artifacts as proposed in the conceptual model, \textbf{it remains unclear how various sources of SATD can be combined} (such as issues and code comments) to facilitate a more complete identification of TD in an automation strategy. Third, \textbf{there is still a need to investigate how the identified SATD items can be utilized in such strategy}. For instance, they could be used in combination with items identified by static analysis (to support the calculation of TD amount), or as a driving force for running static analysis on specific files (to increase the efficiency of the identification process).

Finally, we provided \textbf{ a conceptual model} that describes the current state of TDM Automation as reported in the literature. This model can drive future research and act as a baseline for the development of new approaches (e.g., new toolboxes or automation artifacts). 

\subsection{Implications for Practitioners}
\label{sec:discussion_implications_practitioners}

In \textbf{RQ1}, we provide a \textbf{catalog of automation artifacts} that are available online and ready to be used. We discuss several technical characteristics of the automation artifacts, such as their triggers, input/output and interfaces. Hence, practitioners can use these results to understand the features of the automation artifacts, and subsequently select which artifacts to apply and how. 
In addition, as stated by Silva et al.~\cite{Silva2022}, there is still a lack of tools/technologies that can handle multiple TD types and TDM Activities at the same time. We take a first step in addressing this by discussing how integrating different automation artifacts can be a good direction to increase the automation of TDM; practitioners can follow our recommendations and try, in practice, the integrations that may help them the most in managing TD.

The \textbf{conceptual model} we propose provides several aspects that must be taken into account during the implementation of automation strategies in an industrial setting. Moreover, the model also highlights the limitations of current approaches for TDM automation, which can also help practitioners when adopting current automation solutions (e.g., single automation artifacts or integration between artifacts). 

From a vendors' perspective, \textbf{our study highlights some limitations in the execution of the automation artifacts}, which can help tool developers to improve their automation artifacts. For instance, we found in Section~\ref{sec:results_usage_automation artifacts} that most automation artifacts still demand human intervention to be executed. In this context, we highly recommend vendors to start looking into automating some functionality with development bots; bots are good examples of automation artifacts can run in the background without human intervention, thus improving the level of TDM automation.

\section{Threats to Validity}
\label{sec:tov}

In this section, we discuss the threats to the validity of our SMS. We also discuss the measures we took to mitigate those threats, considering the guidelines provided by Ampatzoglou et al.~\cite{Ampatzoglou2019}. Threats in the context of this study can be classified as Study Selection Validity, Data Validity and Research Validity~\cite{Ampatzoglou2019}.

\textbf{Study Selection}. There is a possibility that some relevant studies were omitted or some irrelevant ones were included, which could harm the completeness and correctness of our results. Ineffective search strings and a non-optimal selection of digital libraries are threats related to study selection. Specifically for our study, we decided to use a single database (Scopus) to perform the initial search, which can be a threat to the validity of our study. To mitigate these threats we followed the guidelines proposed by Wohlin et al.~\cite{Wohlin2022}, which include the search using the Scopus Database followed by forward and backward snowballing. Another threat related to study selection is the possible bias in the selection of the studies. In order to mitigate this one, we conducted a pilot study to calibrate several aspects of our search, such as the search string, the time spam which would be considered, and the inclusion/exclusion criteria. Finally, another possible threat is the decision not to assess the quality of the included primary studies. This decision followed the guidelines proposed by Petersen et al.~\cite{Petersen2015}, which argue against the need of conducting quality assessments in SMS to include ongoing research or non-empirical studies that can provide valuable information (in our case, artifacts). Having said that, since automation artifacts are the main units of analysis of this SMS, we did assess the level of evidence of each artifact, as reported in Section~\ref{sec:results_sw_artifcats}.

\textbf{Data Validity.} \textit{Unverified data extraction} and \textit{author bias} are among the most critical threats to the validity of the data, as these could lead to unreliable results and conclusions. To mitigate these threats, the first and second authors discussed all data items that should be extracted until they reached a consensus about their meaning and content. Moreover, we validated the extraction form by filling it with data from 13 studies during the pilot study. Any disagreement regarding the data items was discussed by the first and second authors, leading to refining and finalizing the extraction form. Finally, during the extraction phase, the data that the first author extracted were revised by the other authors independently.

\textbf{Research validity}. The replicability is one of the main concerns regarding our study, since qualitative and quantitative data analysis and synthesis were required to answer the proposed research questions. To address this, we defined a
detailed protocol (see Section~\ref{sec:design}) based on well-established guidelines~\cite{Kitchenham2015, Wohlin2022,Felizardo2017} for systematic literature reviews. Moreover, we prepared a replication package\footnote{\replicationpkgurl}, which contains all the intermediate outcomes of the process, such as search string calibration, study selection process, and datasets.

\section{Conclusion}
\label{sec:conclusion}

The TDM process encompasses several activities that help practitioners to keep TD under control during the development of software systems. In this study, we reviewed 121 automation artifacts (obtained from 178 studies) that can be applied to automate some of those activities, improving the efficiency of TDM. From the results, a conceptual model was proposed to help practitioners and researchers understand the challenges related to TDM automation that require further investigation. Considering the overall results we found, we argue that more TDM automation should be feasible in the near future. But for that, research is needed to properly and efficiently integrate automation artifacts in software development workflows.

As part of our future work, we intend to validate the proposed conceptual model. In particular, conducting a formal evaluation of the model (such as through expert surveys) could provide insights into its completeness, comprehensibility, and usability. Furthermore, instantiating the model and assessing it within controlled and real-world environments is essential to identify potential areas for improvement. Moreover, we plan to conduct a more in-depth analysis on the usage of automation artifacts in the industry. Specifically, the analysis of how automation artifacts have been evaluated and the challenges in using them from the developers' perspective is still necessary. Additionally, we plan to conduct surveys with developers to gain a deeper understanding of the requirements related to the automation of TDM.

\bibliographystyle{elsarticle-num}
\bibliography{references}

\end{document}